\documentclass{aastex63}
\usepackage{xcolor}
\usepackage{longtable}
\usepackage{amsmath,amssymb}
\usepackage{rotating}
\usepackage{amssymb}
\usepackage{url}

\makeatletter

\newcommand{\Rmnum}[1]{\expandafter\@slowromancap\romannumeral #1@}
\makeatother

\submitjournal{ApJL}

\shorttitle{\emph{Insight}-HXMT observed a huge X-ray flare during the unusual low-luminosity state.}
\shortauthors{Kong et al.}

\begin{document}

\title{\emph{Insight}-HXMT observations of a possible fast transition from jet to wind dominated state during a huge flare of GRS~1915+105}

\author{L. D. Kong\textsuperscript{*}}
\email{kongld@ihep.ac.cn}
\affil{Key Laboratory for Particle Astrophysics, Institute of High Energy Physics, Chinese Academy of Sciences, 19B Yuquan Road, Beijing 100049, China}
\affil{University of Chinese Academy of Sciences, Chinese Academy of Sciences, Beijing 100049, China}

\author{S. Zhang\textsuperscript{*}}
\email{szhang@ihep.ac.cn}
\affil{Key Laboratory for Particle Astrophysics, Institute of High Energy Physics, Chinese Academy of Sciences, 19B Yuquan Road, Beijing 100049, China}

\author{Y. P. Chen\textsuperscript{*}}
\email{chenyp@ihep.ac.cn}
\affil{Key Laboratory for Particle Astrophysics, Institute of High Energy Physics, Chinese Academy of Sciences, 19B Yuquan Road, Beijing 100049, China}

\author{S. N. Zhang}
\affil{Key Laboratory for Particle Astrophysics, Institute of High Energy Physics, Chinese Academy of Sciences, 19B Yuquan Road, Beijing 100049, China}
\affil{University of Chinese Academy of Sciences, Chinese Academy of Sciences, Beijing 100049, China}

\author{L. Ji}
\affil{Institut f{\"u}r Astronomie und Astrophysik, Kepler Center for Astro and Particle Physics, Eberhard Karls, Universit{\"a}t, Sand 1, D-72076 T{\"u}bingen, Germany}

\author{P. J. Wang}
\affil{Key Laboratory for Particle Astrophysics, Institute of High Energy Physics, Chinese Academy of Sciences, 19B Yuquan Road, Beijing 100049, China}
\affil{University of Chinese Academy of Sciences, Chinese Academy of Sciences, Beijing 100049, China}

\author{L. Tao}
\affil{Key Laboratory for Particle Astrophysics, Institute of High Energy Physics, Chinese Academy of Sciences, 19B Yuquan Road, Beijing 100049, China}

\author{M. Y. Ge}
\affil{Key Laboratory for Particle Astrophysics, Institute of High Energy Physics, Chinese Academy of Sciences, 19B Yuquan Road, Beijing 100049, China}

\author{C. Z. Liu}
\affil{Key Laboratory for Particle Astrophysics, Institute of High Energy Physics, Chinese Academy of Sciences, 19B Yuquan Road, Beijing 100049, China}

\author{L. M. Song}
\affil{Key Laboratory for Particle Astrophysics, Institute of High Energy Physics, Chinese Academy of Sciences, 19B Yuquan Road, Beijing 100049, China}
\affil{University of Chinese Academy of Sciences, Chinese Academy of Sciences, Beijing 100049, China}

\author{F. J. Lu}
\affil{Key Laboratory for Particle Astrophysics, Institute of High Energy Physics, Chinese Academy of Sciences, 19B Yuquan Road, Beijing 100049, China}

\author{J. L. Qu}
\affil{Key Laboratory for Particle Astrophysics, Institute of High Energy Physics, Chinese Academy of Sciences, 19B Yuquan Road, Beijing 100049, China}
\affil{University of Chinese Academy of Sciences, Chinese Academy of Sciences, Beijing 100049, China}

\author{T. P. Li}
\affil{Key Laboratory for Particle Astrophysics, Institute of High Energy Physics, Chinese Academy of Sciences, 19B Yuquan Road, Beijing 100049, China}
\affil{Department of Astronomy, Tsinghua University, Beijing 100084, China}
\affil{University of Chinese Academy of Sciences, Chinese Academy of Sciences, Beijing 100049, China}

\author{Y. P. Xu}
\affil{Key Laboratory for Particle Astrophysics, Institute of High Energy Physics, Chinese Academy of Sciences, 19B Yuquan Road, Beijing 100049, China}


\author{X. L. Cao}
\affil{Key Laboratory for Particle Astrophysics, Institute of High Energy Physics, Chinese Academy of Sciences, 19B Yuquan Road, Beijing 100049, China}

\author{Y. Chen}
\affil{Key Laboratory for Particle Astrophysics, Institute of High Energy Physics, Chinese Academy of Sciences, 19B Yuquan Road, Beijing 100049, China}

\author{Q. C. Bu}
\affil{Key Laboratory for Particle Astrophysics, Institute of High Energy Physics, Chinese Academy of Sciences, 19B Yuquan Road, Beijing 100049, China}

\author{C. Cai}
\affil{Key Laboratory for Particle Astrophysics, Institute of High Energy Physics, Chinese Academy of Sciences, 19B Yuquan Road, Beijing 100049, China}

\author{Z. Chang}
\affil{Key Laboratory for Particle Astrophysics, Institute of High Energy Physics, Chinese Academy of Sciences, 19B Yuquan Road, Beijing 100049, China}

\author{G. Chen}
\affil{Key Laboratory for Particle Astrophysics, Institute of High Energy Physics, Chinese Academy of Sciences, 19B Yuquan Road, Beijing 100049, China}

\author{L. Chen}
\affil{Department of Astronomy, Beijing Normal University, Beijing 100088, China}

\author{T. X. Chen}
\affil{Key Laboratory for Particle Astrophysics, Institute of High Energy Physics, Chinese Academy of Sciences, 19B Yuquan Road, Beijing 100049, China}

\author{W. W. Cui}
\affil{Key Laboratory for Particle Astrophysics, Institute of High Energy Physics, Chinese Academy of Sciences, 19B Yuquan Road, Beijing 100049, China}

\author{Y. Y. Du}
\affil{Key Laboratory for Particle Astrophysics, Institute of High Energy Physics, Chinese Academy of Sciences, 19B Yuquan Road, Beijing 100049, China}

\author{G. H. Gao}
\affil{Key Laboratory for Particle Astrophysics, Institute of High Energy Physics, Chinese Academy of Sciences, 19B Yuquan Road, Beijing 100049, China}
\affil{University of Chinese Academy of Sciences, Chinese Academy of Sciences, Beijing 100049, China}

\author{H. Gao}
\affil{Key Laboratory for Particle Astrophysics, Institute of High Energy Physics, Chinese Academy of Sciences, 19B Yuquan Road, Beijing 100049, China}
\affil{University of Chinese Academy of Sciences, Chinese Academy of Sciences, Beijing 100049, China}

\author{M. Gao}
\affil{Key Laboratory for Particle Astrophysics, Institute of High Energy Physics, Chinese Academy of Sciences, 19B Yuquan Road, Beijing 100049, China}

\author{Y. D. Gu}
\affil{Key Laboratory for Particle Astrophysics, Institute of High Energy Physics, Chinese Academy of Sciences, 19B Yuquan Road, Beijing 100049, China}

\author{J. Guan}
\affil{Key Laboratory for Particle Astrophysics, Institute of High Energy Physics, Chinese Academy of Sciences, 19B Yuquan Road, Beijing 100049, China}

\author{C. C. Guo}
\affil{Key Laboratory for Particle Astrophysics, Institute of High Energy Physics, Chinese Academy of Sciences, 19B Yuquan Road, Beijing 100049, China}
\affil{University of Chinese Academy of Sciences, Chinese Academy of Sciences, Beijing 100049, China}

\author{D. W. Han}
\affil{Key Laboratory for Particle Astrophysics, Institute of High Energy Physics, Chinese Academy of Sciences, 19B Yuquan Road, Beijing 100049, China}

\author{Y. Huang}
\affil{Key Laboratory for Particle Astrophysics, Institute of High Energy Physics, Chinese Academy of Sciences, 19B Yuquan Road, Beijing 100049, China}

\author{J. Huo}
\affil{Key Laboratory for Particle Astrophysics, Institute of High Energy Physics, Chinese Academy of Sciences, 19B Yuquan Road, Beijing 100049, China}

\author{S. M. Jia}
\affil{Key Laboratory for Particle Astrophysics, Institute of High Energy Physics, Chinese Academy of Sciences, 19B Yuquan Road, Beijing 100049, China}

\author{W. C. Jiang}
\affil{Key Laboratory for Particle Astrophysics, Institute of High Energy Physics, Chinese Academy of Sciences, 19B Yuquan Road, Beijing 100049, China}

\author{J. Jin}
\affil{Key Laboratory for Particle Astrophysics, Institute of High Energy Physics, Chinese Academy of Sciences, 19B Yuquan Road, Beijing 100049, China}

\author{B. Li}
\affil{Key Laboratory for Particle Astrophysics, Institute of High Energy Physics, Chinese Academy of Sciences, 19B Yuquan Road, Beijing 100049, China}

\author{C. K. Li}
\affil{Key Laboratory for Particle Astrophysics, Institute of High Energy Physics, Chinese Academy of Sciences, 19B Yuquan Road, Beijing 100049, China}

\author{G. Li}
\affil{Key Laboratory for Particle Astrophysics, Institute of High Energy Physics, Chinese Academy of Sciences, 19B Yuquan Road, Beijing 100049, China}

\author{W. Li}
\affil{Key Laboratory for Particle Astrophysics, Institute of High Energy Physics, Chinese Academy of Sciences, 19B Yuquan Road, Beijing 100049, China}

\author{X. Li}
\affil{Key Laboratory for Particle Astrophysics, Institute of High Energy Physics, Chinese Academy of Sciences, 19B Yuquan Road, Beijing 100049, China}

\author{X. B. Li}
\affil{Key Laboratory for Particle Astrophysics, Institute of High Energy Physics, Chinese Academy of Sciences, 19B Yuquan Road, Beijing 100049, China}

\author{X. F. Li}
\affil{Key Laboratory for Particle Astrophysics, Institute of High Energy Physics, Chinese Academy of Sciences, 19B Yuquan Road, Beijing 100049, China}

\author{Z. W. Li}
\affil{Key Laboratory for Particle Astrophysics, Institute of High Energy Physics, Chinese Academy of Sciences, 19B Yuquan Road, Beijing 100049, China}

\author{X. H. Liang}
\affil{Key Laboratory for Particle Astrophysics, Institute of High Energy Physics, Chinese Academy of Sciences, 19B Yuquan Road, Beijing 100049, China}

\author{J. Y. Liao}
\affil{Key Laboratory for Particle Astrophysics, Institute of High Energy Physics, Chinese Academy of Sciences, 19B Yuquan Road, Beijing 100049, China}

\author{B. S. Liu}
\affil{Department of Astronomy, Tsinghua University, Beijing 100084, China}

\author{H. W. Liu}
\affil{Key Laboratory for Particle Astrophysics, Institute of High Energy Physics, Chinese Academy of Sciences, 19B Yuquan Road, Beijing 100049, China}

\author{H. X. Liu}
\affil{Key Laboratory for Particle Astrophysics, Institute of High Energy Physics, Chinese Academy of Sciences, 19B Yuquan Road, Beijing 100049, China}

\author{X. J. Liu}
\affil{Key Laboratory for Particle Astrophysics, Institute of High Energy Physics, Chinese Academy of Sciences, 19B Yuquan Road, Beijing 100049, China}

\author{X. F. Lu}
\affil{Key Laboratory for Particle Astrophysics, Institute of High Energy Physics, Chinese Academy of Sciences, 19B Yuquan Road, Beijing 100049, China}

\author{Q. Luo}
\affil{Key Laboratory for Particle Astrophysics, Institute of High Energy Physics, Chinese Academy of Sciences, 19B Yuquan Road, Beijing 100049, China}
\affil{University of Chinese Academy of Sciences, Chinese Academy of Sciences, Beijing 100049, China}

\author{T. Luo}
\affil{Key Laboratory for Particle Astrophysics, Institute of High Energy Physics, Chinese Academy of Sciences, 19B Yuquan Road, Beijing 100049, China}

\author{R. C. Ma}
\affil{Key Laboratory for Particle Astrophysics, Institute of High Energy Physics, Chinese Academy of Sciences, 19B Yuquan Road, Beijing 100049, China}

\author{X. Ma}
\affil{Key Laboratory for Particle Astrophysics, Institute of High Energy Physics, Chinese Academy of Sciences, 19B Yuquan Road, Beijing 100049, China}

\author{B. Meng}
\affil{Key Laboratory for Particle Astrophysics, Institute of High Energy Physics, Chinese Academy of Sciences, 19B Yuquan Road, Beijing 100049, China}

\author{Y. Nang}
\affil{Key Laboratory for Particle Astrophysics, Institute of High Energy Physics, Chinese Academy of Sciences, 19B Yuquan Road, Beijing 100049, China}
\affil{University of Chinese Academy of Sciences, Chinese Academy of Sciences, Beijing 100049, China}

\author{J. Y. Nie}
\affil{Key Laboratory for Particle Astrophysics, Institute of High Energy Physics, Chinese Academy of Sciences, 19B Yuquan Road, Beijing 100049, China}

\author{G. Ou}
\affil{Key Laboratory for Particle Astrophysics, Institute of High Energy Physics, Chinese Academy of Sciences, 19B Yuquan Road, Beijing 100049, China}

\author{X. Q. Ren}
\affil{Key Laboratory for Particle Astrophysics, Institute of High Energy Physics, Chinese Academy of Sciences, 19B Yuquan Road, Beijing 100049, China}
\affil{University of Chinese Academy of Sciences, Chinese Academy of Sciences, Beijing 100049, China}

\author{N. Sai}
\affil{Key Laboratory for Particle Astrophysics, Institute of High Energy Physics, Chinese Academy of Sciences, 19B Yuquan Road, Beijing 100049, China}
\affil{University of Chinese Academy of Sciences, Chinese Academy of Sciences, Beijing 100049, China}

\author{X. Y. Song}
\affil{Key Laboratory for Particle Astrophysics, Institute of High Energy Physics, Chinese Academy of Sciences, 19B Yuquan Road, Beijing 100049, China}

\author{L. Sun}
\affil{Key Laboratory for Particle Astrophysics, Institute of High Energy Physics, Chinese Academy of Sciences, 19B Yuquan Road, Beijing 100049, China}

\author{Y. Tan}
\affil{Key Laboratory for Particle Astrophysics, Institute of High Energy Physics, Chinese Academy of Sciences, 19B Yuquan Road, Beijing 100049, China}

\author{Y. L. Tuo}
\affil{Key Laboratory for Particle Astrophysics, Institute of High Energy Physics, Chinese Academy of Sciences, 19B Yuquan Road, Beijing 100049, China}
\affil{University of Chinese Academy of Sciences, Chinese Academy of Sciences, Beijing 100049, China}

\author{C. Wang}
\affil{Key Laboratory of Space Astronomy and Technology, National Astronomical Observatories, Chinese Academy of Sciences, Beijing 100012,China}
\affil{University of Chinese Academy of Sciences, Chinese Academy of Sciences, Beijing 100049, China}

\author{L. J. Wang}
\affil{Department of Astronomy, Beijing Normal University, Beijing 100088, China}

\author{W. S. Wang}
\affil{Key Laboratory for Particle Astrophysics, Institute of High Energy Physics, Chinese Academy of Sciences, 19B Yuquan Road, Beijing 100049, China}

\author{Y. S. Wang}
\affil{Key Laboratory for Particle Astrophysics, Institute of High Energy Physics, Chinese Academy of Sciences, 19B Yuquan Road, Beijing 100049, China}

\author{X. Y. Wen}
\affil{Key Laboratory for Particle Astrophysics, Institute of High Energy Physics, Chinese Academy of Sciences, 19B Yuquan Road, Beijing 100049, China}

\author{B. B. Wu}
\affil{Key Laboratory for Particle Astrophysics, Institute of High Energy Physics, Chinese Academy of Sciences, 19B Yuquan Road, Beijing 100049, China}

\author{B. Y. Wu}
\affil{Key Laboratory for Particle Astrophysics, Institute of High Energy Physics, Chinese Academy of Sciences, 19B Yuquan Road, Beijing 100049, China}
\affil{University of Chinese Academy of Sciences, Chinese Academy of Sciences, Beijing 100049, China}

\author{M. Wu}
\affil{Key Laboratory for Particle Astrophysics, Institute of High Energy Physics, Chinese Academy of Sciences, 19B Yuquan Road, Beijing 100049, China}

\author{G. C. Xiao}
\affil{Key Laboratory for Particle Astrophysics, Institute of High Energy Physics, Chinese Academy of Sciences, 19B Yuquan Road, Beijing 100049, China}
\affil{University of Chinese Academy of Sciences, Chinese Academy of Sciences, Beijing 100049, China}

\author{S. Xiao}
\affil{Key Laboratory for Particle Astrophysics, Institute of High Energy Physics, Chinese Academy of Sciences, 19B Yuquan Road, Beijing 100049, China}
\affil{University of Chinese Academy of Sciences, Chinese Academy of Sciences, Beijing 100049, China}

\author{S. L. Xiong}
\affil{Key Laboratory for Particle Astrophysics, Institute of High Energy Physics, Chinese Academy of Sciences, 19B Yuquan Road, Beijing 100049, China}

\author{R. J. Yang}
\affil{College of physics Sciences \& Technology, Hebei University, No. 180 Wusi Dong Road, Lian Chi District, Baoding City, Hebei Province 071002, China}

\author{S. Yang}
\affil{Key Laboratory for Particle Astrophysics, Institute of High Energy Physics, Chinese Academy of Sciences, 19B Yuquan Road, Beijing 100049, China}

\author{Y. J. Yang}
\affil{Key Laboratory for Particle Astrophysics, Institute of High Energy Physics, Chinese Academy of Sciences, 19B Yuquan Road, Beijing 100049, China}

\author{Y. J. Yang}
\affil{Key Laboratory for Particle Astrophysics, Institute of High Energy Physics, Chinese Academy of Sciences, 19B Yuquan Road, Beijing 100049, China}

\author{Q. B. Yi}
\affil{Key Laboratory for Particle Astrophysics, Institute of High Energy Physics, Chinese Academy of Sciences, 19B Yuquan Road, Beijing 100049, China}
\affil{University of Chinese Academy of Sciences, Chinese Academy of Sciences, Beijing 100049, China}

\author{Q. Q. Yin}
\affil{Key Laboratory for Particle Astrophysics, Institute of High Energy Physics, Chinese Academy of Sciences, 19B Yuquan Road, Beijing 100049, China}

\author{Y. You}
\affil{Key Laboratory for Particle Astrophysics, Institute of High Energy Physics, Chinese Academy of Sciences, 19B Yuquan Road, Beijing 100049, China}

\author{F. Zhang}
\affil{Key Laboratory for Particle Astrophysics, Institute of High Energy Physics, Chinese Academy of Sciences, 19B Yuquan Road, Beijing 100049, China}

\author{H. M. Zhang}
\affil{Key Laboratory for Particle Astrophysics, Institute of High Energy Physics, Chinese Academy of Sciences, 19B Yuquan Road, Beijing 100049, China}

\author{J. Zhang}
\affil{Key Laboratory for Particle Astrophysics, Institute of High Energy Physics, Chinese Academy of Sciences, 19B Yuquan Road, Beijing 100049, China}

\author{P. Zhang}
\affil{Key Laboratory for Particle Astrophysics, Institute of High Energy Physics, Chinese Academy of Sciences, 19B Yuquan Road, Beijing 100049, China}

\author{W. C. Zhang}
\affil{Key Laboratory for Particle Astrophysics, Institute of High Energy Physics, Chinese Academy of Sciences, 19B Yuquan Road, Beijing 100049, China}

\author{W. Zhang}
\affil{Key Laboratory for Particle Astrophysics, Institute of High Energy Physics, Chinese Academy of Sciences, 19B Yuquan Road, Beijing 100049, China}
\affil{University of Chinese Academy of Sciences, Chinese Academy of Sciences, Beijing 100049, China}

\author{Y. F. Zhang}
\affil{Key Laboratory for Particle Astrophysics, Institute of High Energy Physics, Chinese Academy of Sciences, 19B Yuquan Road, Beijing 100049, China}

\author{Y. H. Zhang}
\affil{Key Laboratory for Particle Astrophysics, Institute of High Energy Physics, Chinese Academy of Sciences, 19B Yuquan Road, Beijing 100049, China}
\affil{University of Chinese Academy of Sciences, Chinese Academy of Sciences, Beijing 100049, China}

\author{H. S. Zhao}
\affil{Key Laboratory for Particle Astrophysics, Institute of High Energy Physics, Chinese Academy of Sciences, 19B Yuquan Road, Beijing 100049, China}

\author{X. F. Zhao}
\affil{Key Laboratory for Particle Astrophysics, Institute of High Energy Physics, Chinese Academy of Sciences, 19B Yuquan Road, Beijing 100049, China}
\affil{University of Chinese Academy of Sciences, Chinese Academy of Sciences, Beijing 100049, China}

\author{S. J. Zheng}
\affil{Key Laboratory for Particle Astrophysics, Institute of High Energy Physics, Chinese Academy of Sciences, 19B Yuquan Road, Beijing 100049, China}

\author{Y. G. Zheng}
\affil{Key Laboratory for Particle Astrophysics, Institute of High Energy Physics, Chinese Academy of Sciences, 19B Yuquan Road, Beijing 100049, China}
\affil{College of physics Sciences \& Technology, Hebei  University, No. 180 Wusi Dong Road, Lian Chi District, Baoding City, Hebei Province 071002, China}

\author{D. K. Zhou}
\affil{Key Laboratory for Particle Astrophysics, Institute of High Energy Physics, Chinese Academy of Sciences, 19B Yuquan Road, Beijing 100049, China}
\affil{University of Chinese Academy of Sciences, Chinese Academy of Sciences, Beijing 100049, China}

\begin{abstract}
We present the analysis of the brightest flare that was recorded in the \emph{Insight}-HMXT data set, in a broad energy range (2$-$200 keV) from the microquasar GRS~1915+105 during an unusual low-luminosity state.
This flare was detected by \emph{Insight}-HXMT among a series of flares during 2 June 2019 UTC 16:37:06 to 20:11:36, with a 2-200 keV luminosity of 3.4$-$7.27$\times10^{38}$ erg s$^{-1}$.
Basing on the broad-band spectral analysis, we find that the flare spectrum shows different behaviors during bright and faint epochs. 
The spectrum of the flare can be fitted with a model dominated by a power-law component.
Additional components show up in the bright epoch with a hard tail and in the faint epoch with an absorption line $\sim$ 6.78 keV.
The reflection component of the latter is consistent with an inner disk radius $\sim$ 5 times larger than that of the former.
These results on the giant flare during the ``unusual'' low-luminosity state of GRS~1915+105 may suggest that the source experiences a possible fast transition from a jet-dominated state to a wind-dominated state.
We speculate that the evolving accretion disk and the large-scale magnetic field may play important roles in this peculiar huge flare.
\end{abstract}

\section{Introduction}
The microquasar GRS~1915+105 is a black hole (BH) X-ray binary system, characterized with the abundant variabilities well distinguished from the typical ones. It consists of a BH and a K-M \Rmnum{3} companion star (\citealp{2001A&A...373L..37G}),
which was firstly discovered by GRANAT/WATCH all-sky monitor in 
1992 (\citealp{1992IAUC.5590....2C}).
The source has been being monitored thereafter by a series of telescopes like RXTE/ASM, 
Swift/BAT and MAXI/GSC ect., and detected with complex temporal and spectral properties.
Basing on the light curves and colour-colour diagrams (CCDs), \cite{2000A&A...355..271B} and \cite{2002MNRAS.331..745K} classified the variations of its X-ray flux into at least 14 different classes. 
Some of these variability classes are believed to correlate with the limit cycles of accretion and ejection in an unstable disk, and hence the transitions between A, B and C states are defined in \cite{Belloni2000}.
The state connection between GRS~1915+105 and the canonical black-hole binary (\citealp{RM2006}) is not clear.
\cite{2011ApJ...737...69N} presented the first detailed phase-resolved spectral analysis of the $\rho$ class (``heartbeat'' oscillation), which is characterized by regular oscillations with periods of 50–100 s, observed by Chandra and RXTE between the low C state and the high A/B states.
They found that the jet is active at smaller scales during short X-ray hard state with 10\% cycle near the minimum luminosity, and the disk wind at larger scales may lead the fast spectral transitions in different phases. 
From the ``heartbeat'' state, \cite{Zoghbi2016} also found two wind components with low velocities between 500 and 5000 km s$^{-1}$ and probably two more with high velocities reaching 20,000 km s$^{-1}$ ($\sim$ 0.06 c). 
Such an evolution of wind feature may be associated with a bulge, which is born in the inner disk and moving outward as the instability progresses (\citealp{Zoghbi2016}).
\cite{Neilsen2009} argued that the massive winds can affect the disk accretion flow and hence suppress jet formation or even quench jet, since the massive winds are preferentially but not exclusively detected in softer state where jet emission is generally absent or weak (\citealp{Ponti2012}; \citealp{Homan2016}).
The less understood relations among disk, jet and wind as denoted by these results make GRS~1915+105 an appropriate laboratory for further explorations.

GRS~1915+105 remained bright until July 2018.
It entered an extended ``unusually'' low-flux state thereafter, characterized with a lower flux and a harder spectrum (\citealp{Koljonen2020}). 
Such a dramatic dimming has never been seen before and indicates that the source may be approaching its quiescent state.
Different from other black hole binaries, the contemporary monitorings at radio, IR and X-ray bands detected strong variabilities in this ``unusually'' state of GRS 1915+105 (\citealp{Murata2019}; \citealp{Trushkin2019}; \citealp{Koljonen2019}; \citealp{Trushkin2019b}; \citealp{Trushkin2020};
\citealp{Vishal2019}; \citealp{Murata2019}).   
These variabilities at soft and hard X-rays were observed by MAXI/GSC, NICER, Swift/BAT, Astrosat and Konus-Wind (\citealp{2019ATel12818....1S}; \citealp{2019ATel13308....1H}; \citealp{2019ATel12805....1J}; \citealp{2020ATel13478....1T}; \citealp{2020ATel13652....1A}; \citealp{Neilsen2019}).
A dust ring was detected on May 16, 2019 by Swift/XRT which could be the footprint of the strong flare occurred a few days earlier (\citealp{2019ATel12761....1I}).
\cite{Neilsen2019} found a bright flare from NICER observation on 20 May 2019, which lasted for $200$ s and had a peak flux of 3400 cts/s. Their spectral analysis indicated that the flare has a hard continuum that is composed of a very strong and skewed iron line typical to relativistic reflection and a narrow line at $\sim$ 6.4 keV. 
The average spectrum of the flare itself shows a deep iron absorption at $\sim$ 6.65 keV.
The Chandra gratings revealed a highly obscured state that may be relevant to the so-called 'fail wind': a dense, massive accretion disk wind origins near the central engine but can not escape because of low velocity (\citealp{Miller2020}).

\emph{Insight}-HXMT has been monitoring GRS 1915+105 and detected one flare with the longest duration and the largest peak flux with respect to any other flares recorded in \emph{Insight}-HXMT data set during this new state.
We analyze this huge flare in a broad-band energy range (2$-$200 keV) covered by the three main detectors of \emph{Insight}-HXMT: LE, ME and HE. 
In this letter, we introduce the data reduction and analysis in Section 2, and present the results in Section 3. 
Finally, we make discussions in Section 4 and conclusion in Section 5.

\section{Observation and Data reduction}
The Hard X-ray Modulation Telescope, also dubbed as \emph{Insight}-HXMT (\citealp{2014SPIE.9144E..21Z}; \citealp{2020SCPMA..63x9502Z}), was launched on June 15 2017 with a broad energy band (1-250 keV) and a large effective area in high energy range.
\emph{Insight}-HXMT consists of three collimated telescopes: the High Energy X-ray Telescope (HE, 18 cylindrical NaI(Tl)/CsI(Na) phoswich detectors) (\citealp{2020SCPMA..63x9503L}), the Medium Energy X-ray Telescope (ME, 1728 Si-PIN detectors) (\citealp{2020SCPMA..63x9504C}), and the Low Energy X-ray Telescope (LE, Swept Charge Device (SCD)) (\citealp{2020SCPMA..63x9505C}), with collecting-area/energy-range of $\sim$5100 $\rm cm^2$/20-250 keV, $\sim$952 $\rm cm^2$/5-30 keV and $\sim$384 $\rm cm^2$/1-10 keV, and typical Field of View (FoV) of $1.6^{\circ}\times6^{\circ}$, $1^{\circ}\times4^{\circ}$ and $1.1^{\circ}\times5.7^{\circ}$; $5.7^{\circ}\times5.7^{\circ}$ for LE, ME and HE, respectively.

\emph{Insight}-HXMT has observed the microquasar GRS~1915+105 since 15 June 2017 (MJD 57919), which sums up to 115 observations and a total exposure of $\sim$ 2250 ks.
\emph{Insight}-HXMT detected a huge flare during 2 June 2019 UTC 16:37:06 to 2 June 2019 UTC 20:11:36 (MJD 58636.69$-$58636.84), which lasts for about 13 ks. 
We focus on this flare and perform the analyses with \emph{Insight}-HXMT Data Analysis Software (HXMTDAS) v2.02 (\url{http://hxmt.org/software.jhtml}). 
The data are selected under a series of criteria as recommended by the \emph{Insight}-HXMT team: filter for the good time interval (GTI), elevation angle (ELV) larger than $10^{\circ}$; geometric cutoff rigidity (COR) larger than 8 GeV; offset for the point position smaller than $0.04^{\circ}$; data removal with 300 s more coverage of the South Atlantic Anomaly (SAA) passage. 
The energy bands are adopted for energy spectral analysis as 2-10 keV (LE), 10-35 keV (ME) and 27-200 keV (HE), considering the background level and the exposure. 
The backgrounds are estimated with the official tools: LEBKGMAP, MEBKGMAP and HEBKGMAP in version 2.0.9 based on the standard \emph{Insight}-HXMT background models (\citealp{2020arXiv200401432L}; \citealp{2020arXiv200501661L}; \citealp{2020arXiv200306260G}). 
The XSPEC v12.10.1f software package (\citealp{1996ASPC..101...17A}) is used to perform the spectral fitting. 
Uncertainty estimated for each spectral parameter is $90\%$, and a systematic error of 1$\%$ is added. 
The errors of the parameters are computed using MCMC (Markov Chain Monte Carlo) of length 10,000.
\section{Results and Discussion}
\subsection{The flux and color evolution}
\begin{figure}
    \centering\includegraphics[width=0.49\textwidth]{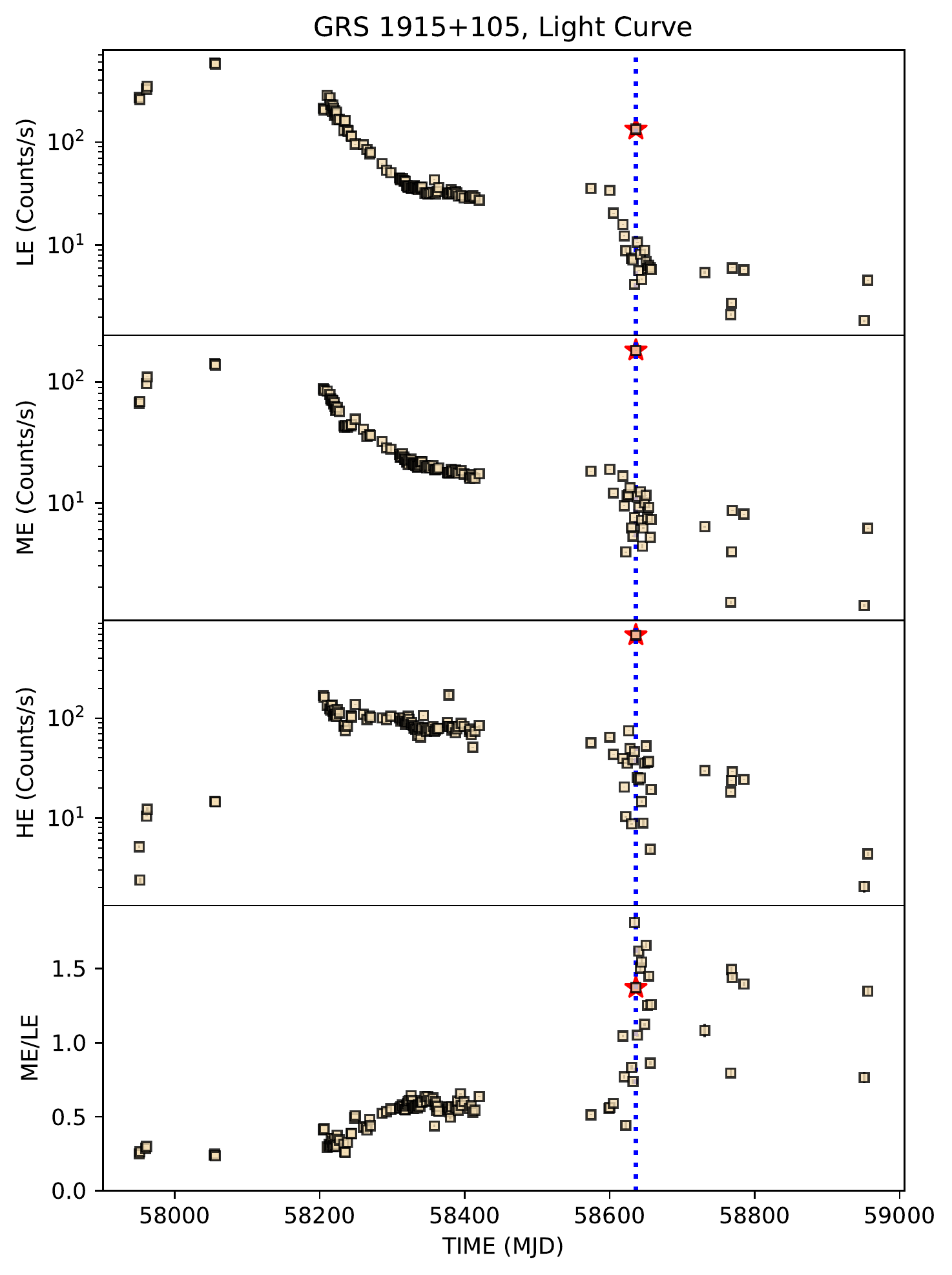}
    \centering\includegraphics[width=0.49\textwidth]{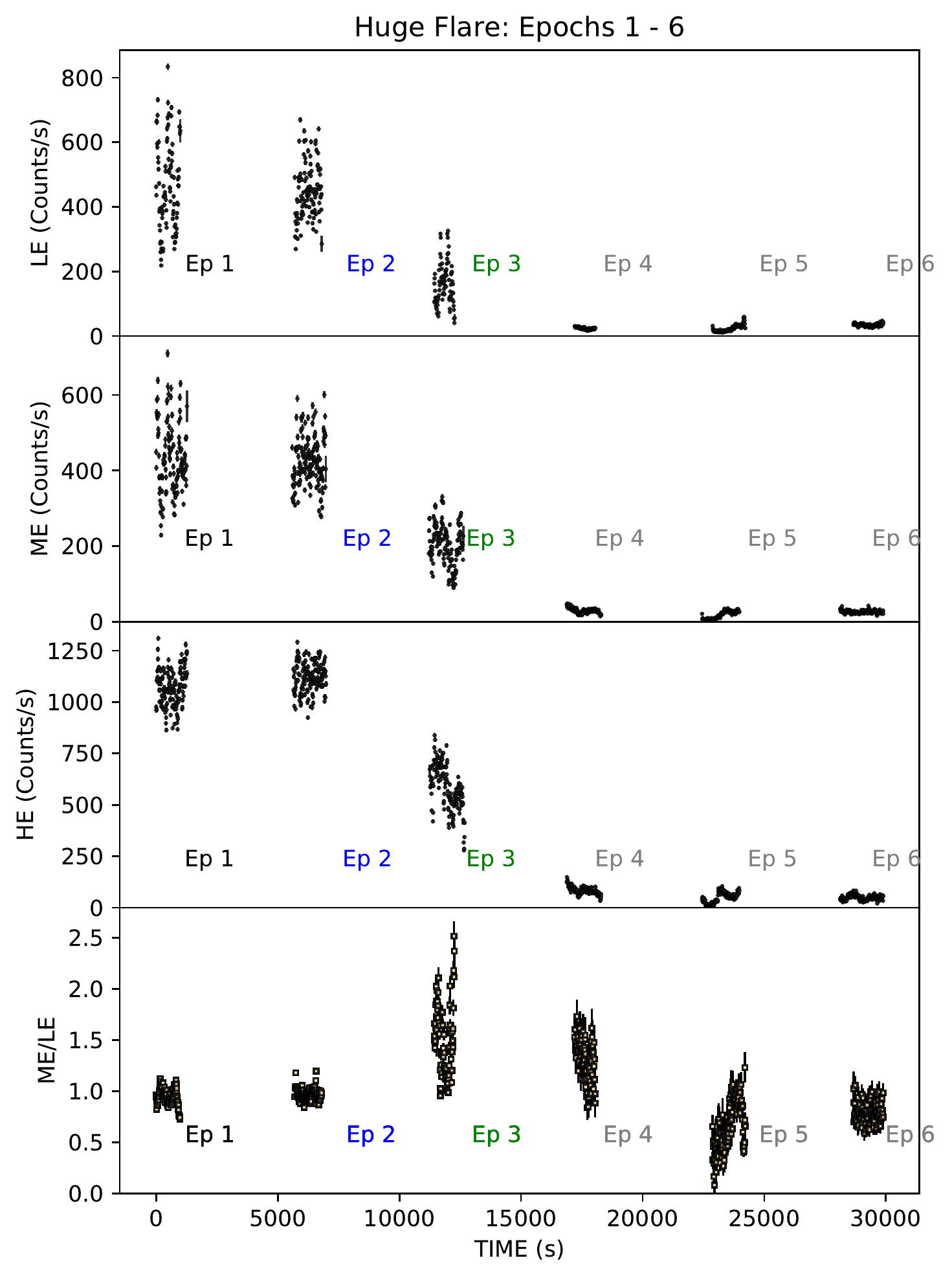}
    \caption{{Left panel: The light curves show observations in a broad-band energy range by LE (1 - 10 keV, top), ME (10 - 20 keV, middle) and HE (30 - 150 keV, bottom) with \emph{Insight}-HXMT from 2017-07-11 to 2020-04-17. The hardness ratio of ME/LE is plotted in the bottom, and it shows significant transition around the blue dotted line representing the position of the huge flare. The red star marks the huge flare.
    Right panel: The detailed light curve and hardness ratio (10 s time bin) of the flare observed by \emph{Insight}-HXMT from 2 June 2019 T 16:37:06 to 2 June 2019 T 20:11:36} (MJD 58636.69 to 58636.84). We seperate the flare into three epochs.
    }
    \label{LC_plot}
\end{figure}

The \emph{Insight}-HXMT observations are shown in the left panel in Figure~\ref{LC_plot} (1 day time-bin for each point), where an overall long-term evolution of the GRS~1915+105 is obvious. 
The top three panels are the count rates of LE (1$-$10 keV), ME (10$-$20 keV) and HE (30$-$150 keV), respectively, and the bottom panel shows the hardness ratio of the ME/LE (soft color band, SC) in 1 day time-bin.
The source evolved to a low hard state at around MJD 58300, with the hardness increasing from $\sim$ 0.25 to 0.6.
Its flux decreased after MJD 58600 with a rapid increase of hardness ratio.
Different from the quiescence of other black hole binaries, there are a forest of X-ray and radio flares in this ``unusual'' low-luminosity state (\citealp{Murata2019}; \citealp{Trushkin2019}; \citealp{Trushkin2019b}; \citealp{Koljonen2019}; \citealp{Koljonen2020}; \citealp{Trushkin2020}).
The brightest X-ray flare observed with \emph{Insight}-HMXT occurred from 2 June 2019 UTC 16:37:06 (MJD 58636.69) to 2 June 2019 UTC 20:11:36 (MJD 58636.84). Following this event, a huge flare at radio band (RATAN-600) appeared during 3 June 2019 at UTC 00:11 (\citealp{Koljonen2019}). 
The right panel of Figure~\ref{LC_plot} shows the details of this flare observed in ObsID: P010131007501: the flare was decaying in snapshots of the first three  (denoted as Epochs 1$-$3) but almost ceased in  the others  (denoted as Epochs 4$-$6).
These monitorings suggest the flare has a duration longer than 13 ks.
Since the source had similar counts rate in Epochs 1 and 2: $\sim$ 400 ct/s for LE and ME, and $\sim$ 1000 ct/s for HE, but droped by half in Epoch 3, our analyses are focused  on the first three snapshots.
    
\begin{figure}
    \centering\includegraphics[scale=0.9]{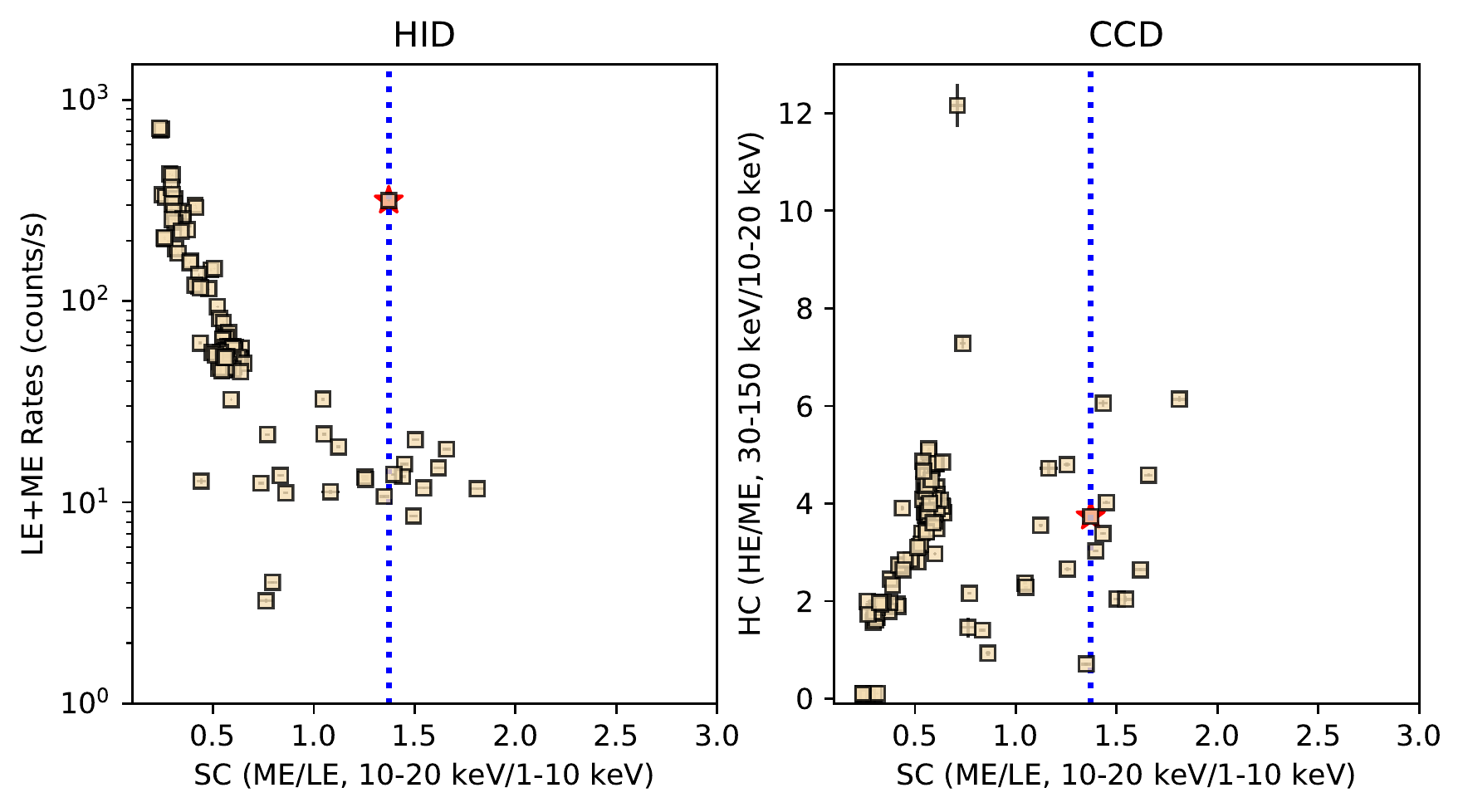}
    \centering\includegraphics[scale=0.9]{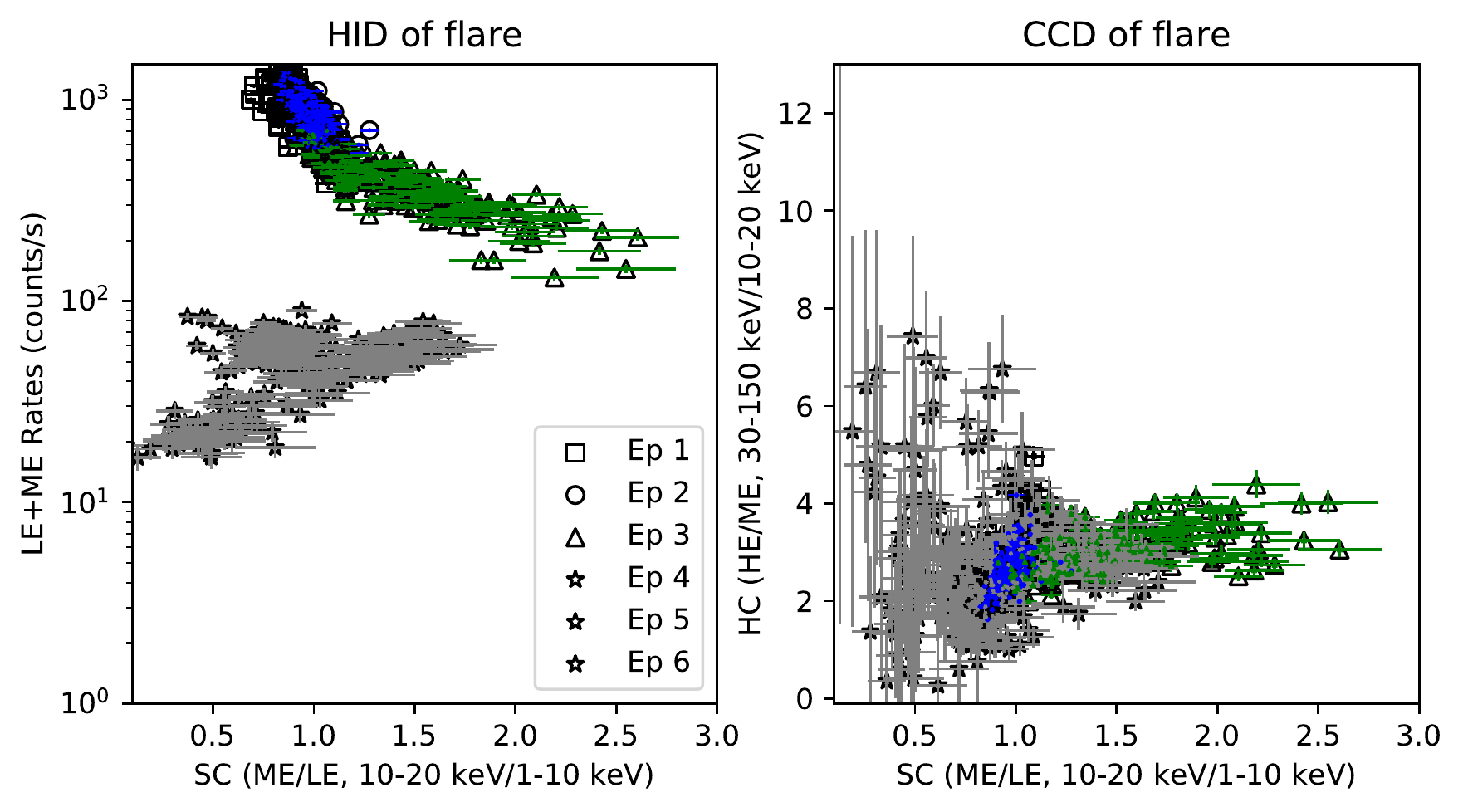}
    \caption{{The CCD and HID of GRS~1915+105 with the same duration of Fig. 2. The blue dotted line shows the position of the huge flare. The red star marks the huge flare. For the bottom panels, we separate the huge flare into six epochs, which is plot in different colors as the same meaning as right panel in Figure 2.}
    }
    \label{HD_plot}
\end{figure}

As for the investigations of the hardness-intensity diagram (HID) and color-color diagram (CCD), we put this flare into the context of long-term evolution as observed by \emph{Insight}-HXMT and find that the flare stands out significantly only in the HID (see the top two panels of Figure~\ref{HD_plot}).   
The CCD and HID of the flare itself (see the lower two panels of Figure~\ref{HD_plot}) show that Epochs 1$-$2 are located in a region significantly different from Epoch 3 and the Epochs 4$-$6. 
This indicates that the source may experience spectral transition during this peculiar short flaring period.

\subsection{The spectral analysis}
\begin{figure}
    \centering\includegraphics[angle=-90, width=0.49\textwidth]{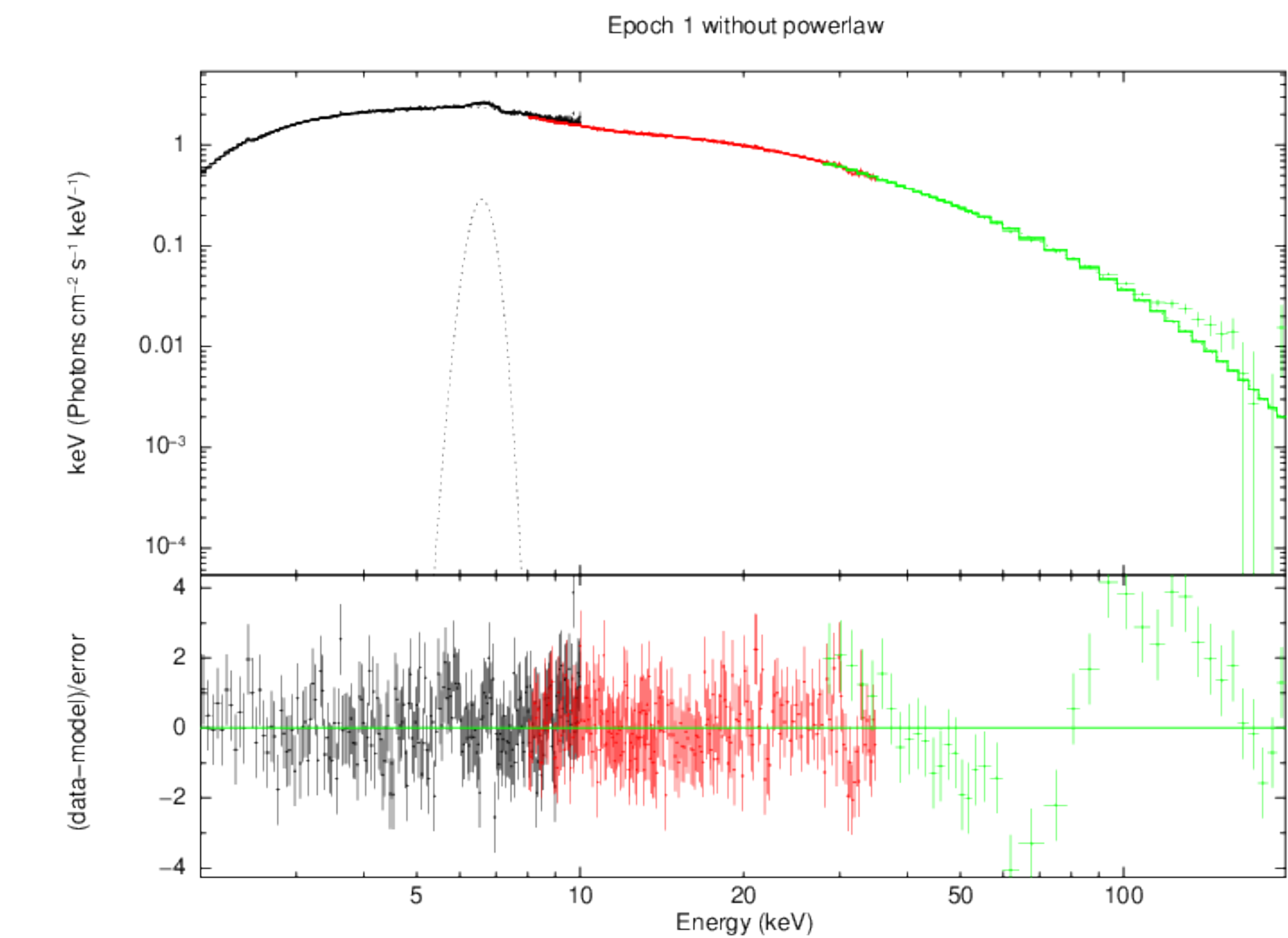}
    \centering\includegraphics[angle=-90, width=0.49\textwidth]{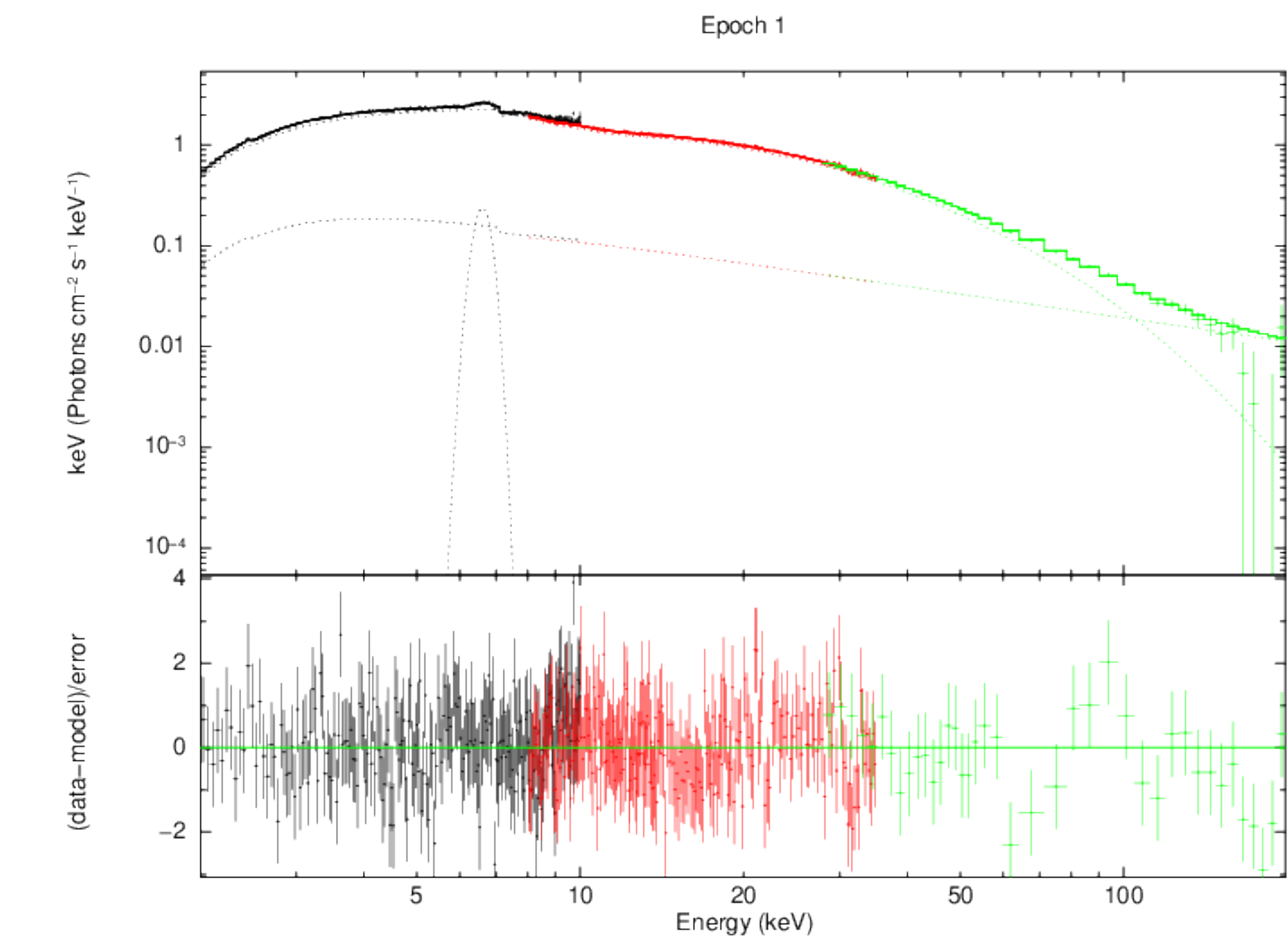}
    \centering\includegraphics[angle=-90, width=0.49\textwidth]{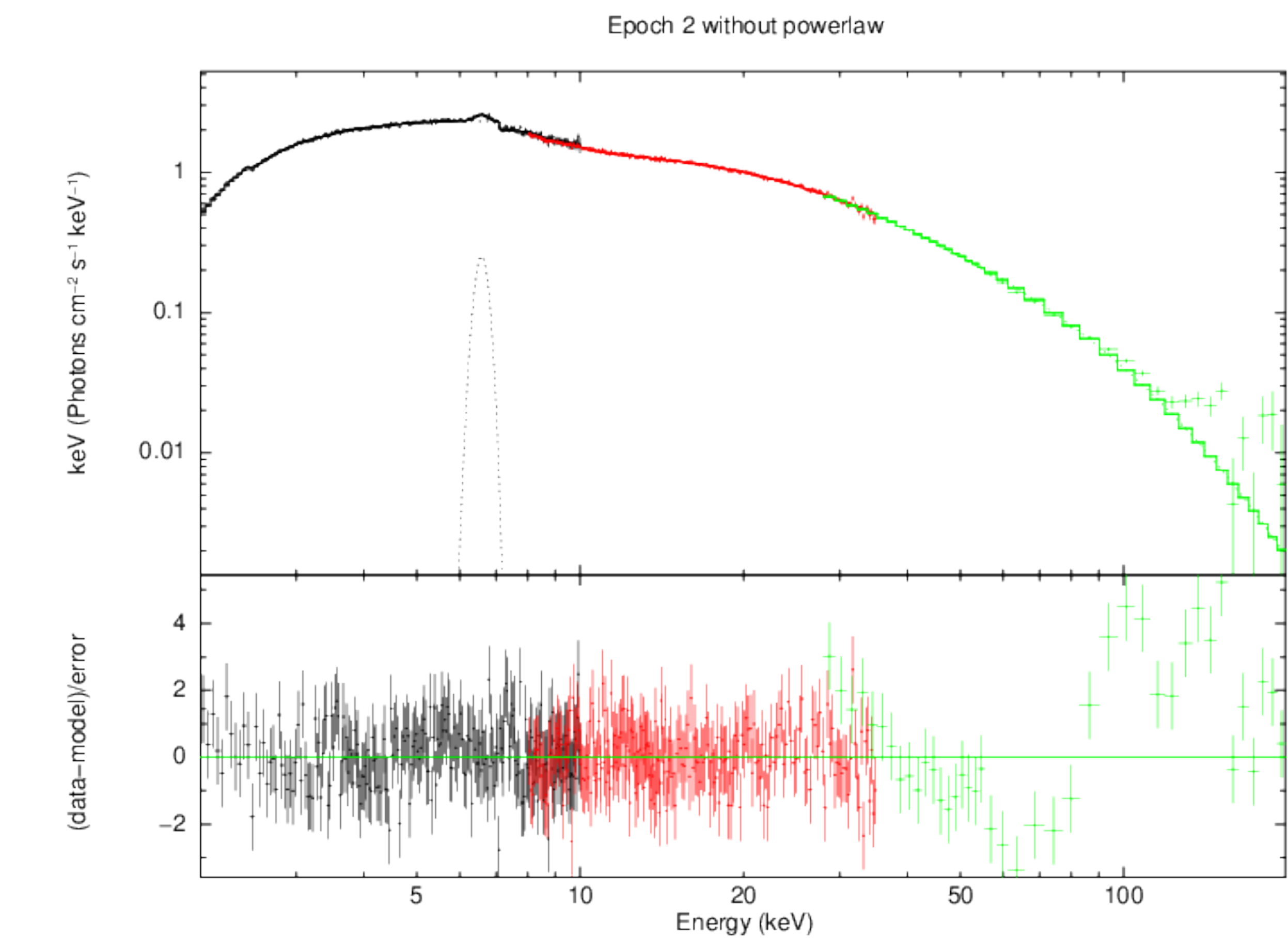}
    \centering\includegraphics[angle=-90, width=0.49\textwidth]{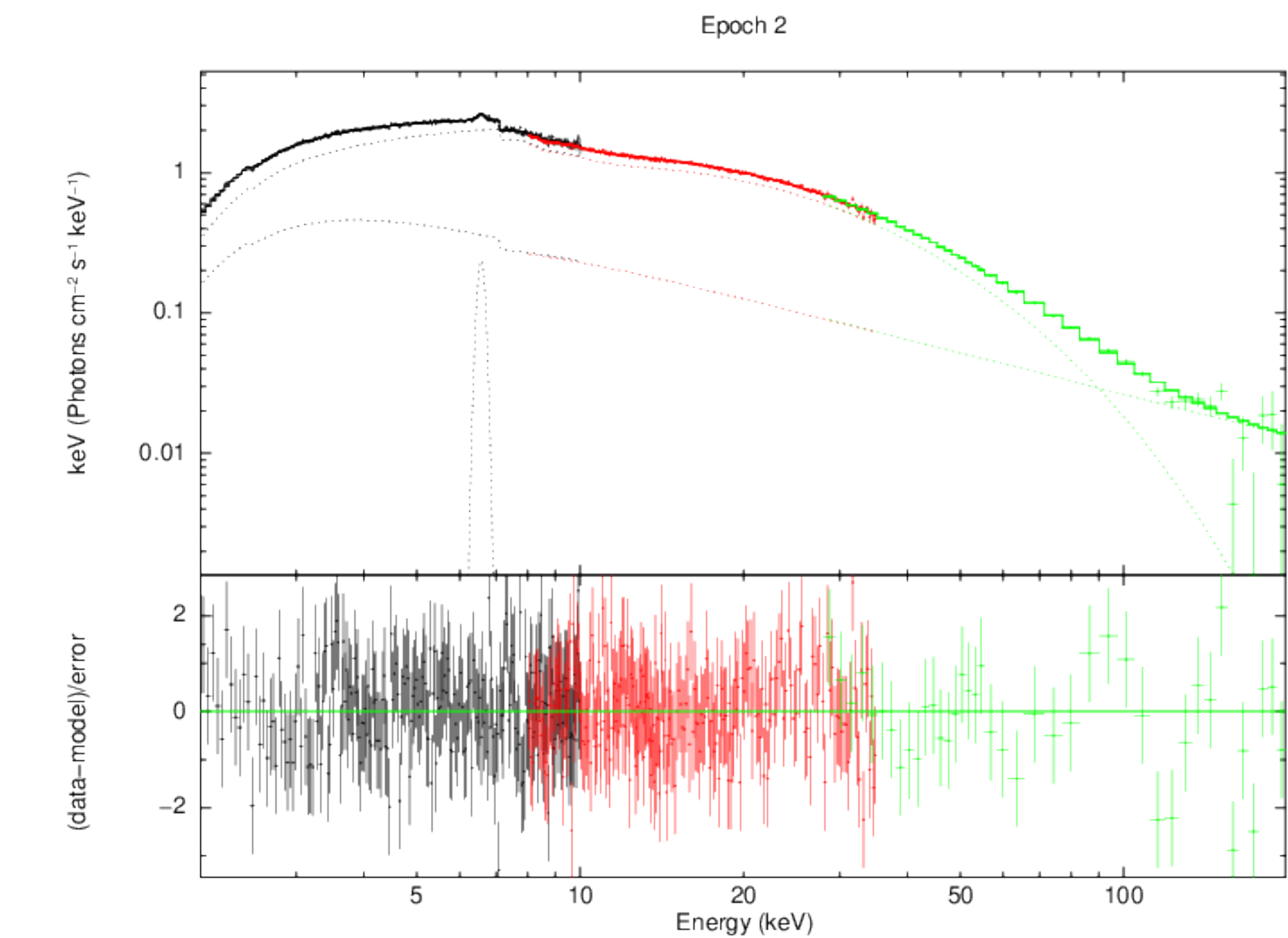}
    \centering\includegraphics[angle=-90, width=0.49\textwidth]{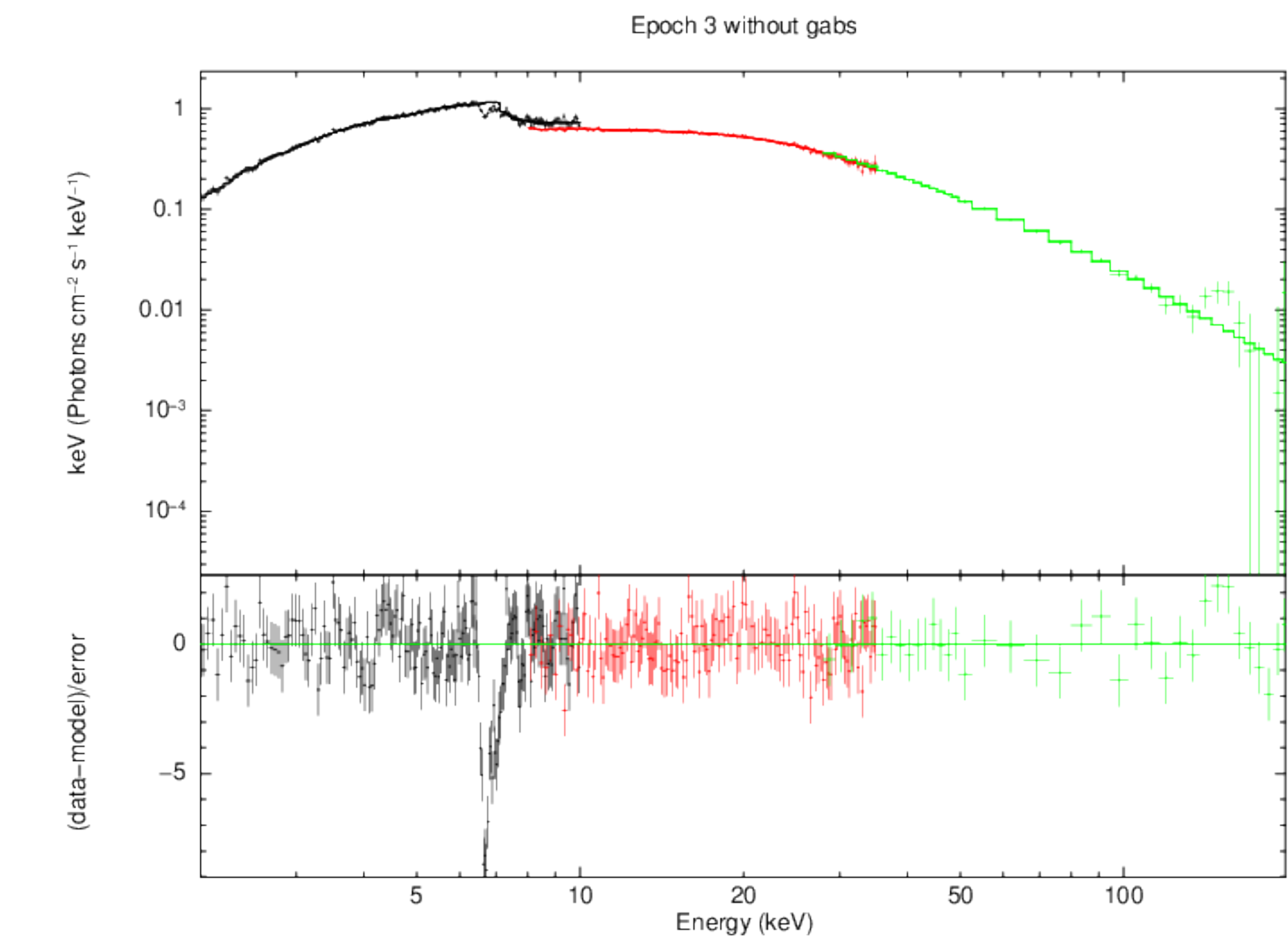}
    \centering\includegraphics[angle=-90, width=0.49\textwidth]{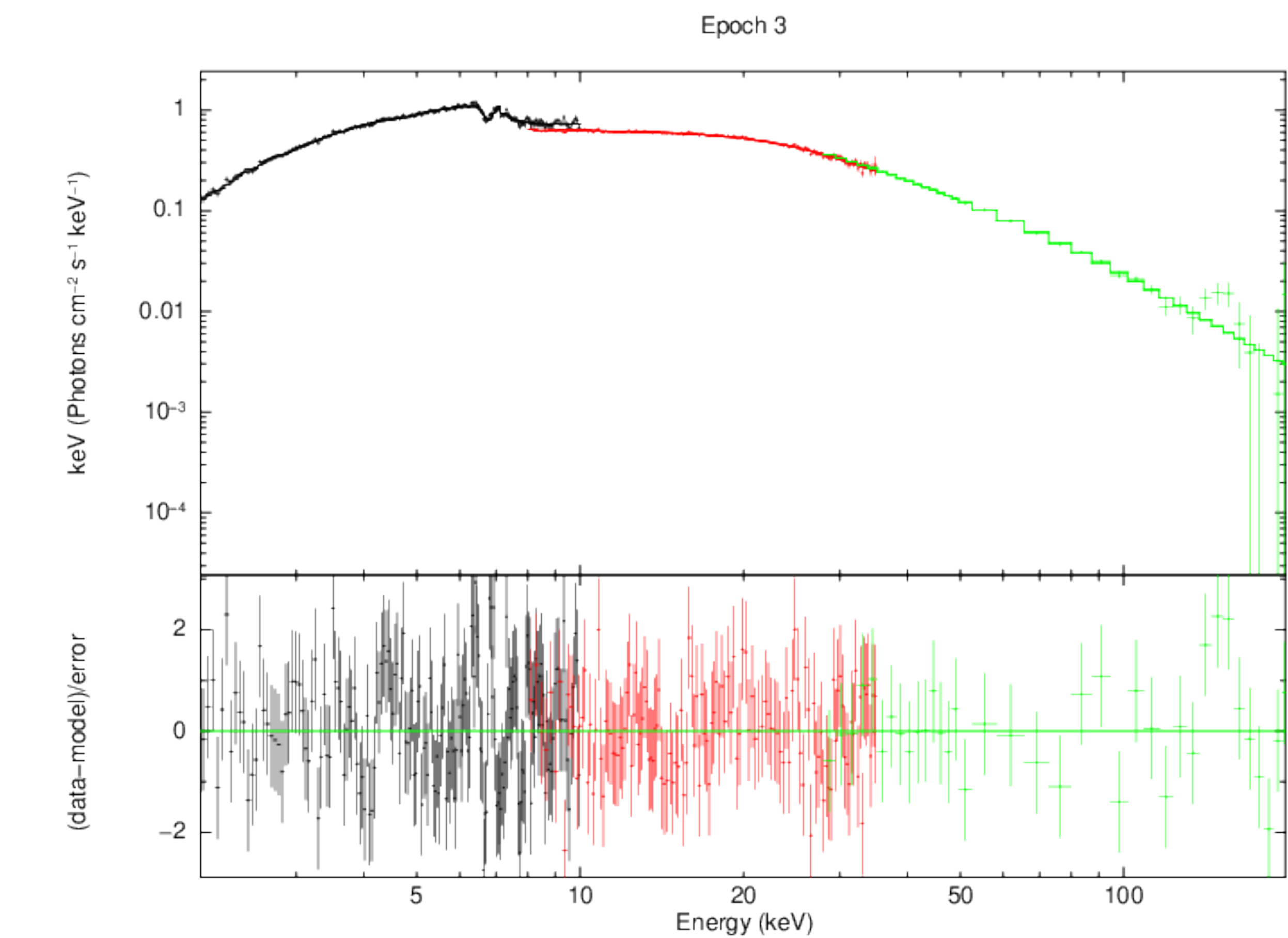}
    \caption{{The pictures show different Epochs during the flare. The LE, ME and HE are noticed as follows: 2 - 10 keV, 10 - 35 keV and 27 - 200 keV, respectively. The parameters and other details of fittings are listed in Table~\ref{spectral_fitting}. For the bottom right panel, we find the $powerlaw$ component is not required.}
    }
    \label{spectra}
\end{figure}

\begin{figure}
    \centering\includegraphics{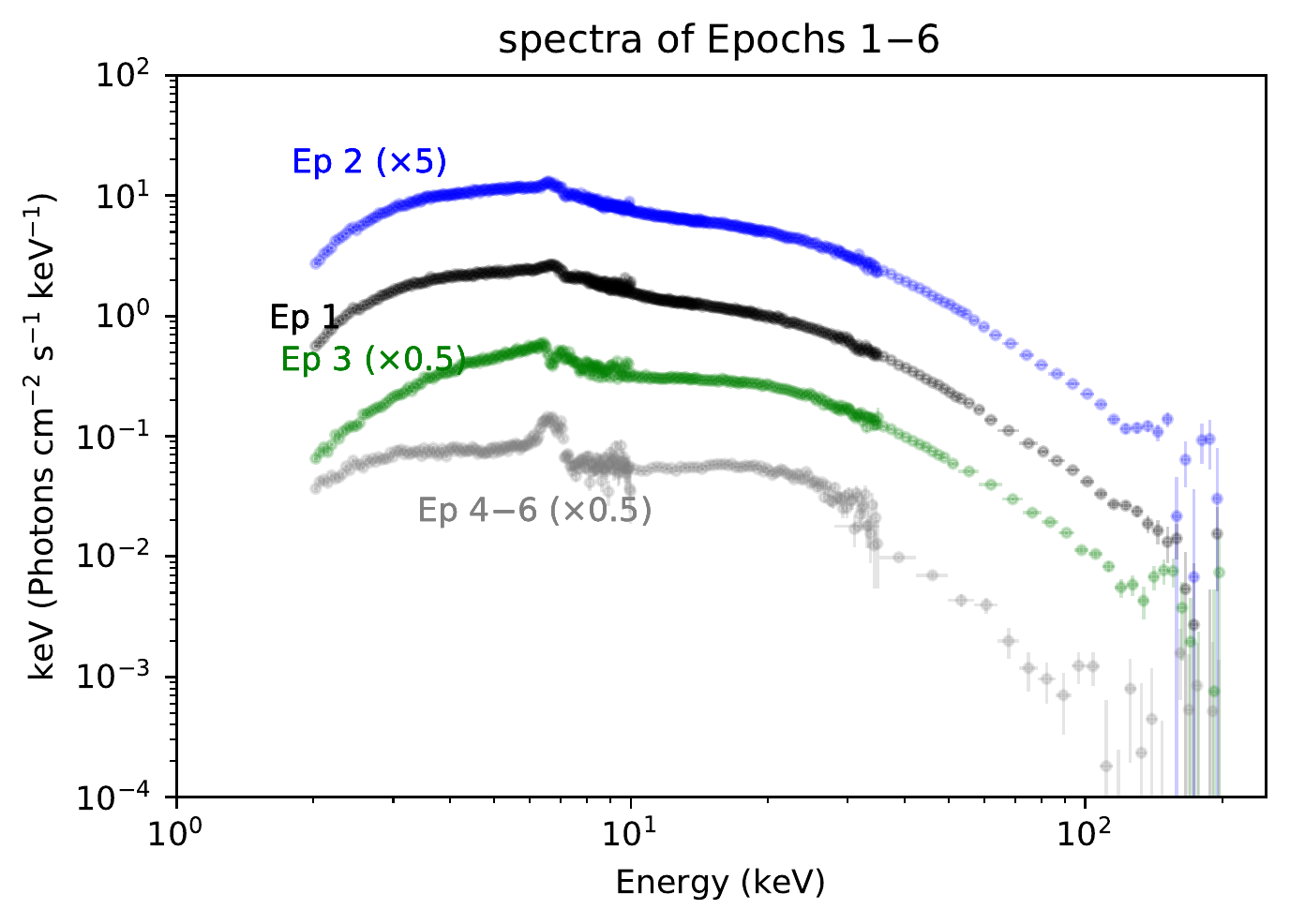}
    \caption{The picture shows the spectra of six epochs of the huge flare. For the Epoch 4$-$6, we combine the spectra for higher significance. The colors of the points follow the Figure~\ref{LC_plot}. The Epochs 1$-$6 show significant evolution.}
    \label{spectra_EP1-6}
\end{figure}

The spectral analyses are focused on the Epochs 1$-$3 (see Table~\ref{spectral_fitting} and Figure~\ref{spectra}) and comparisons are made with respect to Epochs 4$-$6 (see Figure~\ref{spectra_EP1-6}).
For GRS~1915+105, the mass of the black hole,
distance to the source, and inclination angle of the disk are taken as 12.4 $M_{\odot}$, 8.6 kpc, and $60^{\circ}$ in estimating the inner radius and luminosity (\citealp{2014ApJ...796....2R}).
The spectral analysis is carried out in a broad energy range (2$-$200 keV), which covers 2$-$10 keV for LE, 8$-$35 keV for ME and 27$-$200 keV for HE.
The first trial of fitting the Epochs 1$-$3 with $tbabs$$\times$$nthcomp$ results in residuals and $\chi_{\rm red}^{2}$ $>$ 2, are mostly attributed to an obvious asymmetry of the broadened iron line and a Compton hump between 10$-$30 keV.
An absorption edge structure appears around 7.1 keV during Epochs 1 and 2, but not in Epoch 3.
The absorption edge was observed both by NICER and HXMT to accompany with an emission line in the reflection component (\citealp{Neilsen2020ApJ...902..152N}). 
Its origin so far still remains unclear but we speculate it may relate to the disk absorption of both the incident and reflected photons at Fe K edge during the Compton scattering process.
Accordingly, we use \emph{edge} model to improve the fitting during Epochs 1 and 2.
Besides, an additional strong narrow absorption structure around 6.7$-$7 keV presents in Epoch 3. 
Accordingly, the relativistic reflection model $relxill$ (\citealp{Garc_a_2014}; \citealp{2014MNRAS.444L.100D}) for three epochs and a gaussian absorption line model $gabs$ in Epoch 3 are introduced to improve the $\chi_{\rm red}^{2}$.
However, residuals still exist clearly above 50 keV for Epochs 1 and 2 (see upper left and middle left panels in Figure~\ref{spectra}) which required a power-law component.
Finally adoption of a model \emph{tbabs}$\times$\emph{edge}$\times$(\emph{gaussian}+\emph{relxill}+\emph{powerlaw}) can fit the spectra with a $\chi_{\rm red}^{2}$ = 0.99 and 0.96 for Epochs 1 and 2, respectively. 
For Epoch 3 one needs a modified model of \emph{tbabs}$\times$\emph{pcfabs}$\times$\emph{gabs}$\times$\emph{relxill}, which results in $\chi_{\rm red}^{2}$ = 0.96. The $pcfabs$ is used to account for additional absorption circumstance around BH and avoid the sudden rising of the $n_{\rm H}$ with $tbabs$ (\citealp{Koljonen2020}). 

Based on these spectral fittings, the luminosities (2$-$200 keV) are estimated for the each Epoch: $\sim$ 7.24 and 7.27 $\times$ $10^{38}$ erg/s for Epochs 1 and 2 and $\sim$ 3.4 $\times$ $10^{38}$ erg/s for Epoch 3.
The hydrogen column density $N_{\rm H}$ is estimated with the Tuebingen-Boulder ISM absorption model \emph{tbabs} \citep{2000ApJ...542..914W} as 4$-$5$\times$10$^{22}$ cm$^{-2}$ for the three Epochs.
As for Epoch 3, $pcfabs$ model gives an additional absorption of $\sim$ 11.8$\times$10$^{22}$ cm$^{-2}$ from the matter with a covering fraction of $\sim$ 0.63 during Epoch 3.  

For \emph{relxill} components in the three Epochs, the spin $a$ = 0.98 and the inclination angle $\theta$ = $60^{\circ}$ are fixed during the spectral fittings (\citealp{2014ApJ...796....2R}).
In the \emph{relxill} model, the reflect fraction $R_{\rm f}$ is defined as the ratio of the incident photons intensity that illuminates the accretion disc to that observed directly, $\gamma$ and $E_{\rm cut}$ denote the initial spectral index and its deviation energy from a simple power-law shape.
Other parameters in \emph{relxill} provide information about the accretion disk: the inner radius $R_{\rm in}$, ionization of the accretion disk $\xi$ and iron abundance of the material $A_{\rm Fe}$.
During Epochs 1 and 2, the reflection component can be adequately described by reprocessing of a hard spectrum with low energy cutoff ($\gamma$ $\sim$ 1.02, $E_{\rm cut}$ $\sim$ 22$-$26 keV) in an accretion disk with ionization of log $\xi$ $\sim$ 3.4$-$3.6 and iron abundance $A_{\rm Fe}$ $\sim$ 3.58$-$5 (in units of solar abundance).
However, these parameters change significantly in Epoch 3: lower log $\xi$ $\sim$ 1.4 and $A_{\rm Fe}$ $\sim$ 0.61; softer power-law spectrum with spectral index of $\gamma$ $\sim$ 2.42 and higher cutoff energy $E_{\rm cut}$ $\sim$ 195.95 keV.
The reflection fraction $R_{\rm f}$ increases from $\sim$ 0.28$-$0.79 in Epochs 1 and 2 to 2.71 in Epoch 3.

An overall trend is that the inner disk tends to evolve larger through the three Epochs: the inner radius is about 2 $R_{\rm ISCO}$ in Epoch 1, 5 $R_{\rm ISCO}$ in Epoch 2, but increases to 13 $R_{\rm ISCO}$ in Epoch 3. 
But the F-tests show no significant presence of a disk component in all three Epochs. 
For Epochs 1 and 2, we find that the residuals have some structures above 50 keV, which can be significantly accounted for by introducing a \emph{powerlaw} component in the spectra (see the left panels in Figure~\ref{spectra}).
This component can extend to energies above 100 keV.
The contributions of this component to the total luminosity is derived as $2.93_{-0.85}^{+0.93}$ $\times$ $10^{37}$ erg s$^{-1}$ in Epoch 1 ($\sim$ 3 $\sigma$) and $6.25_{-1.46}^{+1.41}$ $\times$ $10^{37}$ erg s$^{-1}$ in Epoch 2 ($\sim$ 4 $\sigma$).
In Epochs 1 and 2, narrow lines are found in spectra with a centroid energy of $\sim$ 6.6 keV and a line width of $\sigma$ $\sim$ 100 to 200 eV, probably associated with He-like Fe \Rmnum{25} emission.
During Epoch 3, accompanied with the absences of the power law and the emission line, the appearance of an additional absorption line, we find a significant narrow absorption line at $6.78_{-0.03}^{+0.02}$ keV, with a width of $\sigma$ = $143_{-39}^{+26}$ eV, and a strength of $0.13_{-0.01}^{+0.02}$ ($\sim$ 9 $\sigma$, see bottom the panels in Figure~\ref{spectra}).
Considering an energy resolution 140 eV@5.9 keV for LE (\citealp{2020SCPMA..63x9505C}), and just one single absorption line is detected by \emph{Insight}-HXMT, the velocity of the outflow is hard to be precisely estimated. Since the 6.78 keV absorption line may be relevant to the contamination of 6.98 keV, if any, which can not be resolved by LE, we perform a spectral simulation by setting two absorption lines at 6.7 and 6.98 keV with same line width of 0.14 keV but different line depth of 2:1. 
The fitting result shows an absorption line can exist at energy similar to that observed in Epoch 3.
Actually, iron emission lines $\sim$ 6.6 keV are detected once combining the spectra of Epochs 4$-$6 when the source flare almost ceased.

\section{Discussions}
We have carried out detailed analyses on a huge flare of GRS~1915+105 observed by \emph{Insight}-HXMT during the low hard state of the source and obtained a few interesting results during decay phase in Epochs 1$-$3 and ceasing phase in Epochs 4$-$6.

\begin{figure}
    \centering\includegraphics[width=0.80\textwidth]{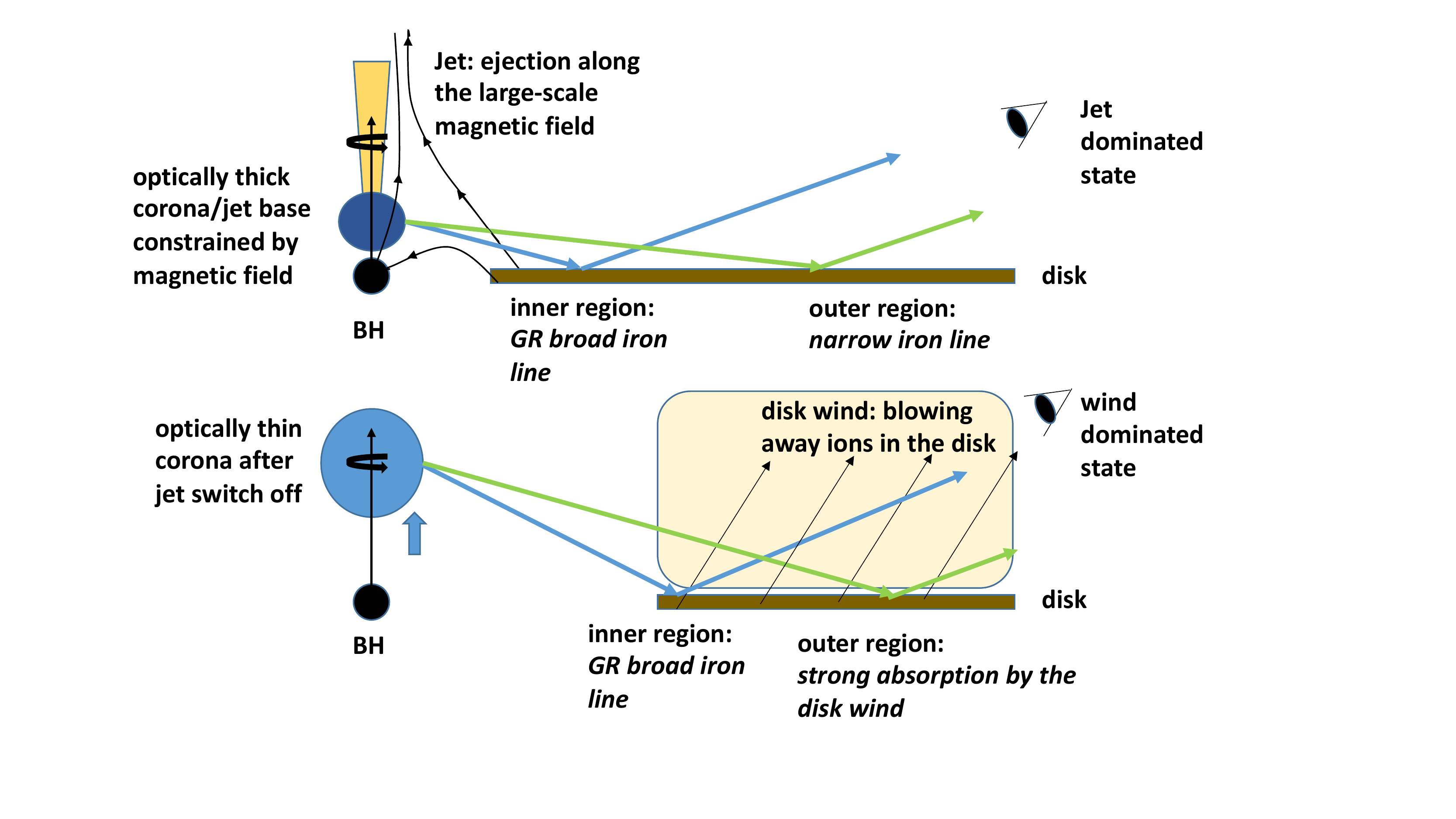}
    \caption{A simplified sketch of the two states during the huge flare. The upper panel illustrates the jet dominated state with large-scale collimated magnetic filed threading the accretion disk near $R_{\rm ISCO}$ and the matters eject along the black arrow lines (filed lines). The optically thick corona/jet base is constraned near the black hole, which irradiates the inner and outer region of the accretion disk, where the broad and the narrow iron lines are produced, respectively. The bottom panel shows the circumstance during the wind-dominated state. The inner radius of the accretion disk moves outward, and the collimated magnetic field turns into diverging field anchor the outer region of the accretion disk. The corona expands to a larger scale and optically thin after the jet switch off. The reflection component shows different properties because of the outflow absorption. The iron absorption line implies magnetic field driven wind or stable absorber on the disk.}
    \label{Toy_model}
\end{figure}

The components of power-law and Fe emission line are present in Epochs 1 and 2, but disappear in Epoch 3, where instead an absorption line shows up at around 6.78 keV. 
The reflection component of the source spectrum suggests that the disk evolves outwards, and the incident spectrum becomes softer and has a higher cutoff energy.
Similar results can be derived as well by introducing instead a reflection model of relxilllp (see Table~\ref{spectral_fitting}).
Both the ionization of the accretion disk and iron abundance of the material decrease but the reflection fraction increases. 
In Epoch 3, the column density is about a factor of a few larger than that in Epochs 1 and 2.  
These results may be understood in a scenario of jet/wind transitions where the disk magnetic field plays an important role (see Figure~\ref{Toy_model}). 

Although the origin of the observed huge flare in GRS 1915+105 is still largely unclear, it can in principle  be driven by either the viscous instabilities in the accretion disk or the magnetic field. 
The former is most likely relevant to the canonical relation of the wind/jet to the spectral states.  
The activity observed in radio band at time around X-ray flares 
may imply that at least part of the flare can be related 
to the jet ejection (\citealp{Murata2019}; \citealp{Trushkin2019}; \citealp{Koljonen2019}; \citealp{Trushkin2019b}; \citealp{Trushkin2020}).

In our results shown in Figure~\ref{spectra}, the significant residuals in Epoch 1 and 2 above 50 keV can be regarded as a ``hard tail'' with a power-law index $\sim$ 2, which may be associated with relativistic jet formed with a large-scale collimated magnetic field near the black hole (\citealp{Reig2016}). 
Meanwhile, the radio activities and flares occurred around the X-ray flares also suggest the correlation of the X-ray flare with relativistic jet (\citealp{Trushkin2020}; \citealp{Mirabel1994}).
On the other hand, such a ``hard tail'' weakens when the flare luminosity becomes lower in Epoch 3 and, instead of having a gaussian emission line in Epochs 1 and 2, an absorption line shows up at $\sim$ 6.78 keV. 
We would like to note that, the limited energy resolution of LE does not allow to constrain the velocity of the wind and hence the possible presence of a `failed' disk wind with low velocity can not be ruled out.
Nevertheless, such a wind feature, as discussed in \cite{Miller2020}, is most likely driven by magnetic field.

Theoretically, the formation of large-scale magnetic fields near a black hole is usually believed to highly relate to the advection of the accretion flow, such as accretion disk and corona. 
The large-scale magnetic field can be dragged inward efficiently by the corona above the disc, the so-called ``coronal mechanism'' \citep{2009ApJ...707..428B}, which provides a way to solve the difficulty of field advection in a geometrically thin accretion disc.
But the maximum power of the jet accelerated by the magnetic field advected by the corona is less than 0.05 Eddington luminosity \citep{2018MNRAS.473.4268C}.
\cite{Li_2019} suggested that the external field can be efficiently dragged inward in a thin disk with magnetic outflows to form the large-scale magnetic field which plays a main role in acceleration and collimation of jet.
Consequently, the disk structure will be altered in the presence of outflows, especially for strong outflows, which can carry both mass and angular momentum from the disk.
Based on their model, the disk may work as a bridge to connect the jet and wind during the flare in case that a magnetic field is at work.
Our results show different disk properties in its inner radius according to the switch off/on of the jet and disk wind during the flare, although the disk emissions are probably too weak to be detected directly.

Jet is usually observed in absence of the presence of disk wind. 
By revealing a surprisingly simple jet-quenching mechanism in GRS 1915+105, \cite{Neilsen2009} pointed out fundamental new insights into the long-term disk-jet coupling around accreting black holes and left attractive evidence of the mechanism by which stellar-mass black holes can regulate their own growth. 
The disk wind in the soft X-ray states can be so powerful in carrying away so much mass that matter is halted from flowing into the jet. 
Our results from the flaring GRS~1915+105 show wind presents at flux lower than jet during Epoch 3. The absence of power law component and significant increasing of $N_{\rm H}$ may imply that the Epoch 3 is intrinsically softer rather than a typical hard state. The detection of the transition between jet and wind can be broadly consistent with \cite{Ponti2012}.
Using an ionization parameter $\rm log\ \xi$ $\sim$ 2.5$-$4, an electron density $n_{\rm e}$ $\sim$ $10^{14}\ \rm cm^{-3}$ from the \cite{Neilsen2020ApJ...902..152N} and a luminosity $L=L_{2-200}$ $\sim$ $3.8\times10^{38}\ \rm erg\ s^{-1}$ during Epoch 3, a launch radius of the wind can be estimated as $R_{\rm Launch}=1.96-10.96\times10^{10} \rm cm$ according to the following formula (\citealp{Tarter1969})
\begin{equation}
R_{\rm Launch} = (L\times(n_{\rm e}\xi)^{-1})^{1/2}.
\end{equation}
Meanwhile, the Compton temperature $T_{\rm IC}$ and Compton radius $R_{\rm C}$ can be estimated as $0.67\times10^{7}\ \rm K$ and $1.23\times10^{12}\ \rm cm$, respectively. 
\cite{Dubus2019} considered the contribution from radiation pressure at luminosity close to Eddington (\citealp{Proga2002}; \citealp{Done2018MNRAS.473..838D}). Accordingly, the equation for $T_{\rm IC}$ and $R_{\rm IC}$ is shown as follow:
\begin{equation}
\frac{T_{\rm IC}}{10^{7}\ \rm K}=\begin{cases}4.2-4.6\times \rm log(l/0.02)\ \rm if\ \emph{l} < 0.02 \\0.36\times(l/0.02)^{1/4}\ \rm if\ \emph{l} \ge 0.02\end{cases},
\end{equation}
\begin{equation}
R_{\rm IC}\approx10^{12}\times\frac{M}{10\ M_{\odot}}\times\frac{10^{7}\ \rm K}{T_{\rm IC}}\times(1-\sqrt{2}\frac{L}{L_{\rm edd}}),
\end{equation}
where $l=L/L_{\rm edd}$.
Since the $R_{\rm Launch}$ is far less than $0.2R_{\rm IC}$, the absorption wind in Epoch 3 is not likely thermal driven.
Thus suggests that a large scale magnetic field may be occasionally at work for boring both jet and disk wind, and responsible for their mutual transitions.

In this ``unusual'' low-luminosity state of GRS~1915+105 a relatively weaker flare was detected by NICER.
The flare reported in this work has a longer duration and a higher luminosity than the flare reported by NICER (\citealp{Neilsen2019}; \citealp{Neilsen2020ApJ...902..152N}).
Our spectral results in Epoch 3 are roughly consistent with those derived at peak of the NICER flare (\citealp{Neilsen2019}; \citealp{Neilsen2020ApJ...902..152N})).
We note that the source flux in Epoch 3 turns out to be comparable to that of the peak flux on the NICER flare. 
\cite{Miller2020} investigated the highly obscured state in the ``unusual'' low-luminosity period, and ended up with a ``failed wind''.
They argued that the wind could not achieve the local escape velocity, and thus leads to a heavily obscured local environment.
It seems that the canonically prohibited disk wind in the hard state can be born but with immaturity, known as ``failed wind'', in presence of the magnetic field. 
Thus ``successful'' disk wind could be the consequence of having stronger strong magnetic field, which may be the case as observed in the huge flare of GRS~1915+105.

In summary, we speculate that the scenario illustrated in Figure~\ref{Toy_model} may account for the spectral evolution as observed in the flare decay from Epochs 1 and 2 to Epoch 3.  
In Epochs 1 and 2 (the upper panel of Figure~\ref{Toy_model}), a jet is formed from channeling the accretion material from the disk via large-scale magnetic field. 
The emissions from the jet base are reflected at the different parts of the disk, where at the inner part broadened Fe line is formed and at the outer part the regular Fe emission lines produced.
Once the flare enters Epoch 3, as shown in the bottom panel of Figure~\ref{Toy_model}, the inner disk recedes from the central black hole and the large-scale magnetic field that connects the former jet and disk accretion material breaks.
A further investigation of the reflection component in spectral fitting with \emph{relxilllp} shows the height of the central illuminating source can largely increase in Epoch 3 (see Table~\ref{spectral_fitting}).  
Also a similar trend of having a receding disk in Epoch 3 is present with \emph{relxilllp}.
The height of the corona can increase significantly from $\sim$ 6 $R_{\rm g}$ during Epochs 1 and 2 to $\sim$ 30 $R_{\rm g}$ during Epoch 3.
Although the errors of $R_{\rm in}$ are relatively large, an overall trend of having a receding inner disk is also visible with \emph{relxillp}.

The constraint upon the geometry of illuminating source may be responsible for the difference of the two reflection models in measurement of the disk radius.
In both models the power law component shows up in Epochs 1 and 2 but disappears in Epoch 3. This gives the evidence of jet evolving fainter in Epoch 3. Also, the larger height of illuminating source in Epoch 3 may suggest that either a higher corona or a faint jet merges into corona which results in larger emission region (see Figure~\ref{Toy_model}).
The open field lines accelerate and channel part of the accretion materials into disk wind, part of which are the ionized Fe.
The interception of the central light house by such disk wind can result in Fe absorption line and an additional absorption with fairly large coverage. 
In Epoch 3, the larger spectral index and higher cutoff energy of the central light house may suggest that the collapsed jet is merged into corona and hence forms a central hot region with a relatively larger height but smaller optical depth with respect to those in Epochs 1 and 2.
Consequently, we observe a larger reflection fraction. 
As for the drop in Fe abundance and Fe emission line, most probably it is due to the absorption by the disk wind, which may lead to decrement of Fe line production at the outer disk.  

\section{Conclusions}
In this study, we have analyzed so far the largest X-ray flare of the microquasar GRS 1915+105 recorded in \emph{Insight}-HXMT data set during its ``unusual'' low-luminosity state. 
The joint diagnostics of this flare with \emph{Insight}-HXMT in a rather broad energy band and a context of contemporary observations in other wavelengths reveals a peculiar flaring source behavior, which may suggest that occasionally the black hole X-ray binary activity may under the control of the presence of a large scale magnetic field and ends up with behavior different from the canonical ones.

\acknowledgments
This work made use of data from the \emph{Insight}-HXMT mission, a project funded by China National Space Administration (CNSA) and the Chinese Academy of Sciences (CAS).
This work is supported by the National Key R\&D Program of China (2016YFA0400800) and the National Natural Science Foundation of China under grants U1838201, 11473027, U1838202, 11733009, and U1838104, U1938101.

\bibliography{ref}
\bibliographystyle{aasjournal}

\begin{table}[ptbptbptb]
\caption{Spectral parameters of three Epochs during the flare}
    \begin{tabular}{cccccccccc}
\hline
\hline
Component & Parameters & Epoch 1 & Epoch 1 & Epoch 2 & Epoch 2 & Epoch 3 & Epoch 3
\\
\hline
tbabs & $n_{\rm H}\ (10^{22}\ \rm cm^{-2})$ & $4.40_{-0.06}^{+0.02}$ & $4.51_{-0.08}^{+0.04}$ & $4.18_{-0.25}^{+0.26}$ & $4.45_{-0.01}^{+0.02}$ & $5.6_{-0.9}^{+0.5}$ & $5.53_{-0.78}^{+0.69}$
\\
\hline
edge & $E_{\rm edge}$ (keV) & 7.1 (fixed) & 7.1 (fixed) & 7.1 (fixed) & 7.1 (fixed) & ... & ...
\\
& $\tau_{\rm Max}$ & $0.07_{-0.02}^{+0.01}$ & $0.03_{-0.01}^{+0.01}$ & $0.10_{-0.01}^{+0.01}$ & $0.10_{-0.01}^{+0.01}$ & ... & ...
\\
\hline
pcfabs & $n_{\rm H}\ (10^{22}\ \rm cm^{-2})$ & ... & ... & ... & ... & $11.8_{-2.0}^{+2.4}$ & $12.0_{-1.3}^{+1.7}$
\\
& $Cvr_{\rm F}$ & ... & ... & ... & ... & $0.63_{-0.07}^{+0.10}$ & $0.64_{-0.08}^{+0.07}$
\\
F-test & $F$-value & ... & ... & ... & ... & 34.6 & ...
\\
& $P$-value & ... & ... & ... & ... & $2.14\times10^{-15}$ & ...
\\
\hline
gaussian & $E_{\rm line}$ (keV) & $6.60_{-0.04}^{+0.05}$ & $6.60_{-0.01}^{+0.02}$ & $6.57_{-0.04}^{+0.03}$ & $6.58_{-0.03}^{+0.03}$ & ... & ...
\\
& $\sigma$ (eV) & $222_{-11}^{+26}$ & $238_{-32}^{+38}$ & $105_{-33}^{+42}$ & $132_{-13}^{+15}$ & ... & ...
\\
& norm ($10^{-2}$) & $2.2_{-0.2}^{+0.5}$ & $2.6_{-0.2}^{+0.3}$ & $1.0_{-0.3}^{+0.2}$ & $1.3_{-0.1}^{+0.1}$ & ... & ...
\\
F-test & $F$-value & 32.6 & ... & 13.7 & ... & ... & ...
\\
& $P$-value & $2.2\times10^{-20}$ & ... & $8.6\times10^{-9}$ & ... & ... & ...
\\
\hline
gabs & $E_{\rm line}$ (keV) & ... & ... & ... & ... & $6.78_{-0.03}^{+0.02}$ & $6.75_{-0.01}^{+0.01}$
\\
& $\sigma$ (eV) & ... & ... & ... & ... & $143_{-39}^{+26}$ & $119_{-30}^{+20}$
\\
& strength & ... & ... & ... & ... & $0.13_{-0.01}^{+0.02}$ & $0.13_{-0.01}^{+0.01}$
\\
\hline
powerlaw & $\Gamma$ & $1.78_{-0.05}^{+0.03}$ & $1.86_{-0.06}^{+0.02}$ & $1.98_{-0.02}^{+0.02}$ & $1.48_{-0.05}^{+0.05}$ & ... & ...
\\
& norm & $0.75_{-0.22}^{+0.12}$ & $1.07_{-0.20}^{+0.12}$ & $2.47_{-0.30}^{+0.23}$ & $0.16_{-0.04}^{+0.04}$ & ... & ...
\\
F-test & $F$-value & 73.29 & ... & 71.63 & ... & ... & ...
\\
& $P$-value & $4.2\times10^{-31}$ & ... & $1.93\times10^{-30}$ & ... & ... & ...
\\
\hline
relxill & $R_{\rm in}$ ($R_{\rm ISCO}$) & $2.2_{-0.3}^{+0.9}$ & ... & $5_{-1}^{+1}$ & ... & $11_{-3}^{+1}$ & ...
\\
& $\gamma$ & $1.04_{-0.02}^{+0.02}$ & ... & $1.00_{-0.01}^{+0.02}$ & ... & $2.42_{-0.02}^{+0.04}$ & ...
\\
& $a$ & 0.98(fixed) & ... & 0.98(fixed) & ... & 0.98 (fixed) & ...
\\
& $\theta$ & $60^{\circ}$ (fixed) & ... & $60^{\circ}$ (fixed) & ... & $60^{\circ}$ (fixed) & ...
\\
& log $\xi$ & $3.60_{-0.02}^{+0.11}$ & ... & $3.44_{-0.04}^{+0.02}$ & ... & $1.40_{-0.03}^{+0.03}$ & ...
\\
& $A_{\rm Fe}$ & $3.58_{-0.17}^{+0.64}$ & ... & $\leq5.00$ & ... & $0.61_{-0.05}^{+0.04}$ & ...
\\
& $E_{\rm cut}$ (keV) & $26_{-1}^{+1}$ & ... & $22.33_{-0.23}^{+0.31}$ & ... & $196_{-12}^{+19}$ & ...
\\
& $R_{\rm f}$ & $0.79_{-0.04}^{+0.09}$ & ... & $0.28_{-0.02}^{+0.03}$ & ... & $2.71_{-0.08}^{+0.38}$ & ...
\\
& $N_{\rm rel}$ ($10^{-2}$) & $3.70_{-0.17}^{+0.25}$ & ... & $5.93_{-0.26}^{+0.32}$ & ... & $8.31_{-0.63}^{+0.56}$ & ...
\\
\hline
relxilllp & $h$ ($GM/c^{2}$) & ... & $5.56_{-0.63}^{+0.52}$ & ... & $6.78_{-0.32}^{+0.27}$ & ... & $30_{-4}^{+4}$
\\
& $R_{\rm in}$ ($R_{\rm ISCO}$) & ... & $1.34_{-0.04}^{+0.04}$ & ... & $1.14_{-0.04}^{+0.03}$ & ... & $5_{-3}^{+3}$
\\
& $\gamma$ & ... & $1.06_{-0.02}^{+0.02}$ & ... & $1.11_{-0.01}^{+0.01}$ & ... & $2.43_{-0.05}^{+0.02}$
\\
& log $\xi$ & ... & $3.67_{-0.03}^{+0.05}$ & ... & $3.57_{-0.03}^{+0.03}$ & ... & $1.40_{-0.03}^{+0.03}$
\\
& $A_{\rm Fe}$ & ... & $3.33_{-0.12}^{+0.15}$ & ... & $3.26_{-0.19}^{+0.15}$ & ... & $0.60_{-0.07}^{+0.05}$
\\
& $E_{\rm cut}$ (keV) & ... & $23.59_{-0.68}^{+0.76}$ & ... & $25.70_{-0.43}^{+0.26}$ & ... & $196_{-34}^{+15}$
\\
& $R_{\rm f}$ & ... & $4.51_{-0.47}^{+0.75}$ & ... & $3.94_{-0.29}^{+0.27}$ & ... & $10.00_{-0.01}^{+0.01}$
\\
& $N_{\rm rel}$ ($10^{-2}$) & ... & $5.93_{-0.92}^{+0.35}$ & ... & $5.83_{-0.27}^{+0.29}$ & ... & $9.87_{-0.61}^{+0.79}$
\\
\hline
$L_{\rm PL}$ ($10^{37}$ erg/s) & 2$-$200 keV & $2.93_{-0.85}^{+0.93}$ & ... & $6_{-1}^{+1}$ & ... & ... & ...
\\
\hline
$L_{\rm total}$ ($10^{38}$ erg/s) & 2$-$200 keV & $7.24_{-0.03}^{+0.01}$ & ... & $7.27_{-0.03}^{+0.01}$ & ... & $3.40_{-0.02}^{+0.01}$ & ...
\\
\hline
Fitting & $\chi_{\rm red}^{2}/d.o.f$ & 0.99/1506 & 0.99/1505 & 0.96/1506 & 0.96/1505 & 0.96/1437 & 0.95/1436
\\
\hline
\hline
    \end{tabular}
    \label{spectral_fitting}
\begin{list}{}{}
    \item[Note]{: Uncertainties are reported at the 90\% confidence interval and were computed using MCMC (Markov Chain Monte Carlo) of length 10,000. The 1\% system error for LE, ME and HE has been added during sepctral fittings. The fixReflFrac in \emph{relxilllp} is fixed at 0 during the spectral fitting.}
\end{list}
\end{table}

\end{document}